\documentclass[preprint2]{raa}            

\usepackage{graphicx,times}             
\usepackage{natbib}
\usepackage{amssymb,amsmath}
\bibpunct{(}{)}{;}{a}{}{,}

\usepackage[pagebackref=true]{hyperref}

\begin{document}

  \title{Solitary dwarf galaxy groups as tracers of dark matter halos in the local Universe}

   \volnopage{Vol.0 (202x) No.0, 000--000}      
   \setcounter{page}{1}          

   \author{Z. S. Yuan
      \inst{1,2,3}
   \and Z. L. Wen
      \inst{1,2,3}
   \and J. L. Han
      \inst{1,2,3,*}\footnotetext{$*$Corresponding Authors.}\footnotetext{Z. S. Yuan: https:/orcid.org/0000-0002-0009-5949;} \footnotetext{Z. L. Wen: https:/orcid.org/0000-0001-9550-0929;} \footnotetext{J. L. Han: https:/orcid.org/0000-0002-9274-3092.}
   }

   \institute{National Astronomical Observatories, Chinese Academy of Sciences, 20A Datun Road, Chaoyang District, Beijing 100101, China; {\it hjl@nao.cas.cn}\\
        \and
             State Key Laboratory of Radio Astronomy and Technology, Beijing 100101, China\\
        \and
             School of Astronomy, University of Chinese Academy of Sciences, Beijing 100049, China\\
\vs\no
   {\small Received 202x month day; accepted 202x month day}}

\abstract{In $\Lambda$CDM cosmology, galaxies and clusters form within
  dark matter halos and merge in the hierarchical assembly paradigm to
  form massive systems.  Using the released optical survey data, we
  searched for groups composed solely of dwarf galaxies, each with a
  stellar mass $M_*<10^{9.5}~M_{\odot}$. We identified 14 dwarf galaxy
  groups with at least 5 dwarf galaxies, all located within a
  projected radius of 200~kpc and with a line-of-sight velocity of
  $\pm$300~km~s$^{-1}$. We checked photometric and imaging data and
  found that these 14 dwarf galaxy groups are solitary, with no
  neighboring massive galaxies with $M_*>10^{10}~M_{\odot}$ within
  500~kpc and within $\pm$1200~km~s$^{-1}$.  These dwarf galaxies are
  gravitationally bound within halos with a dynamical mass of around
  $M_{\rm dyn} \sim 10^{12}~M_{\odot}$ and a virial radius of less
  than 400~kpc.  The stellar mass fractions of these dwarf galaxy
  groups with $M_{\rm dyn}>10^{12}~M_{\odot}$ are one magnitude lower
  than predicted by the canonical stellar mass and halo mass
  relation. These dwarf galaxy groups, therefore, are indications of
  dark matter halos in the local Universe that host only a few newly
  formed dwarf galaxies.
}

   \authorrunning{Z. S. Yuan, Z. L. Wen \& J. L. Han }            
   \titlerunning{Dwarf galaxy groups as tracers of dark matter halos }  

   \maketitle

%
%
\section{Introduction}           
\label{sect:intro}

The formation and evolution of large-scale cosmic structures are key
topics in modern astronomy.  Within the hierarchical assembly
paradigm, massive systems are formed through the continuous merging
and aggregation of low-mass progenitor halos \citep[e.g.,][]{swj+05,
  k10, bde+19}. In the standard $\Lambda$ Cold Dark Matter
($\Lambda$CDM) cosmology, galaxies are embedded within dark matter
halos \citep[e.g.,][]{wr78}. Simulations show that low-mass halos with
masses near $\sim$$10^{12}M_{\odot}$ often host abundant
substructures, which promotes the formation of groups consisting
purely of dwarf galaxies \citep[hereafter dwarf galaxy groups,
  e.g.,][]{dkm+08,dwg14}. Dwarf galaxies, as the most abundant and
faint galaxy population in the Universe \citep[e.g.,][]{m12,fcm+20},
are ideal tracers for testing the $\Lambda$CDM framework at the poorly
constrained low-mass regime \citep[e.g.,][]{fw12}.

Statistical analyses show that about 5\% of dwarf galaxies have dwarf
companions, and dense dwarf systems with four or more members are rare
in the local universe \citep{bps+18}. Observations of such dwarf
systems remain extremely limited. Previously, only a few dwarf galaxy
groups have been identified: seven dwarf galaxy groups with 3 to 5
members were reported by \citet{slj+17}, plus one system reported by
\citet{psy+24}. Solitary dwarf galaxy groups are valuable for
cosmological and extragalactic studies because they provide a clean
environment for studying dwarf-dwarf interactions, star formation, and
chemical enrichment. They may finally merge to form intermediate-mass
galaxies.

Theoretical models predict baryon suppression in low-mass halos, where
stellar feedback and inefficient gas cooling lower the
baryon-to-dark-matter mass ratio \citep{sd15}. Observationally,
measurements of stellar mass fractions for dwarf-scale halos are still
limited by the small number of known isolated dwarf galaxy
groups. Moreover, the evolutionary properties of dwarf galaxies, such
as their star formation rates and metallicities, are closely affected
by their surrounding environments \citep[e.g.,][]{plk+10}. For
example, the Magellanic Clouds are strongly affected by tidal forces
from the Milky Way. Therefore, they cannot reflect the intrinsic
evolution of isolated dwarf galaxy groups
\citep[e.g.,][]{pgs+98,kva06,kvb+13,cdf+19}. Many previous studies
have focused on field and cluster dwarf galaxies. Dwarf galaxies in
clusters usually experience environmental quenching, tidal stripping,
and metal enrichment, whereas field dwarfs maintain ongoing star
formation at relatively low metallicities \citep[e.g.,][]{bg08}.  The
stellar mass fraction of dark matter halos is an important
observational constraint on galaxy formation models
\citep[e.g.,][]{bcw10,lgb+12}.

Although the Sloan Digital Sky Survey (SDSS) has obtained over two
million extragalactic spectra \citep[e.g.,][]{aaa+23}, it is not
sensitive enough to low-surface-brightness dwarf galaxies. Incomplete
spectroscopic data have long limited the search for dwarf galaxy
groups. The first Data Release \citep[DR1,][]{daa+26} of the Dark
Energy Spectroscopic Instrument survey \citep[DESI, e.g.,][]{daa+22}
improves this situation. It provides over 13 million spectra with
greater depth and completeness, enabling a systematic search for faint
and isolated dwarf galaxy groups. In this work, we use spectroscopic
data from DESI DR1, supplemented by data from the SDSS and the Two
Micron All-Sky Survey \citep[2MASS,][]{hmm+12}, and also the NED
database\footnote{https://ned.ipac.caltech.edu/,} to identify 14 new
gravitationally bound dwarf galaxy groups with at least five members.
All systems are isolated from massive galaxies. We also analyze the
global properties of dwarf galaxy groups and compare them with massive
galaxy clusters.  Throughout this paper, a ﬂat $\Lambda$ Cold Dark
Matter ($\Lambda$CDM) cosmology with $H_{0} = 70{\rm~km~s^{-1}
  Mpc^{-1}}$, $\Omega_{m}=0.3$, and $\Omega_{\Lambda} = 0.7$ is
adopted.


\section{Identification of dwarf galaxy groups}
\label{sect2}

\subsection{Data}
\label{data}

We obtained the photometric and spectroscopic parameters of galaxies
by cross-matching several major catalogs.  The photometric data of
galaxies were taken from the DR9 and DR10 of the DESI Legacy Survey,
with point-source magnitude depths of $m_g=24.0$~mag, $m_r=23.5$ mag,
and $m_z=22.9$ mag, respectively \citep{dsl+19}. Spectroscopic
redshifts of galaxies were mainly taken from the DESI DR1
\citep{daa+26}. The DESI spectroscopy observations did not repeatedly
observe some bright objects, and therefore the spectroscopic data from
the SDSS DR18 \citep{aaa+23}, the 2MASS Redshift Survey
\citep{hmm+12}, and the NED database were supplemented for objects in
the DESI survey area. Spectra taken from the DESI survey mainly cover
the galaxies with $m_{r}<19.5$~mag \citep{daa+26}. By comparing our
combined spectroscopic sample of galaxies with the spectroscopic data
in the Galaxy And Mass Assembly (GAMA) field of $\sim$250 square
degrees \citep{dbr+22}, we found that the completeness of our combined
sample is about 90\% complete at $m_r = 18$~mag and $\sim83\%$ at
$m_r=19.5$~mag.

Stellar masses were calculated for all galaxies following the
empirical relation from \citet{wh24}, which was obtained from the
stellar mass from $z$- and $W1$-band luminosity and $r-z$ color using
the COSMOS2015 catalog. Our mass estimates show good consistency with
the SDSS measurements from MPA-JHU
(https://wwwmpa.mpa-garching.mpg.de/SDSS/DR7/), with a scatter of only
0.1~dex for galaxies of $z<0.1$.  Motivated by the low redshifts of
known dwarf galaxy groups \citep[$z<0.05$,][]{slj+17,psy+24}, we
restricted our search to galaxies with spectroscopic redshifts
$z<0.1$. Duplicate spectroscopic observations of some galaxies can
introduce biases into group identification. We therefore removed
redundant measurements of these individual galaxies if $\Delta
z\le0.001$ and a projected separation $l<10$~kpc, and kept only the
entry with the largest stellar mass. The final parent sample includes
839,827 galaxies, among which 643,750 of the spectroscopic data come
from DESI DR1, 171,548 from SDSS DR18, 6,196 from 2MASS, and the
remaining 18,333 from NED.

\begin{table}
\caption{Basic parameters of dwarf galaxy groups.}
\label{tab1} 
\small
\centering
\setlength{\tabcolsep}{1pt}
\begin{tabular}{lrrcrrccccccc}
\hline\hline             
\multicolumn{1}{c}{Group} & \multicolumn{1}{c}{RA} & \multicolumn{1}{c}{DEC} & $z$ & N$_g$ & \multicolumn{1}{c}{$\delta$} & ${\rm log}_{10}M_{\rm *,tot}$ & \multicolumn{1}{c}{$C_{M_*}$} & ${\rm log}_{10}M_{\rm b}$ & ${\rm log}_{10}M_{\rm dyn}$
& $\sigma_{\rm 3D}$ & $v_{\rm esc,vir}$ & $r_{\rm vir}$\\
\multicolumn{1}{c}{name} & \multicolumn{1}{c}{(J2000)} & \multicolumn{1}{c}{(J2000)} & & & & ($M_{\odot}$) & & ($M_{\odot}$) & ($M_{\odot}$) & ($\rm km/s$)  & ($\rm km/s$) & (kpc)\\
\multicolumn{1}{c}{(1)} & \multicolumn{1}{c}{(2)} & \multicolumn{1}{c}{(3)} & (4) & (5) & (6) & (7) & (8) & (9) & (10) & (11) & (12) & (13)\\
\hline
 J0226$+$30 & 36.59251 & 30.83102 & 0.0170 & 5 & 7.1  & 9.21 &1.00 & 11.82 & 12.33 & 169 & 310 & 335\\
 J0426$-$04* & 66.62140 & -4.67019 & 0.0113 & 5 & 6.0  & 8.05 &0.74 & 10.37 & 10.82 &  39 &  93 & 104\\
 J0948$-$03* &147.07855 & -3.73392 & 0.0130 & 6 &13.4  & 9.41 &0.99 & 11.35 & 11.71 & 107 & 188 & 206\\
 J1054$+$17 &163.70267 & 17.62122 & 0.0039 &11 &12.3  & 9.87 &1.00 & 11.80 & 12.29 & 163 & 295 & 322\\
 J1137$+$58 &174.26193 & 58.41549 & 0.0039 & 9 & 4.8  & 9.47 &1.00 & 12.02 & 12.46 & 219 & 350 & 381\\
 J1140$+$60 &175.12971 & 60.29888 & 0.0041 & 5 & 4.3  & 9.92 &0.99 & 11.07 & 11.60 &  83 & 170 & 188\\
 J1153$-$03 &178.41873 & -3.99639 & 0.0053 & 8 & 6.9  & 9.42 &0.98 & 11.57 & 12.05 & 118 & 246 & 269\\
 J1215$-$00 &183.83398 & -0.39807 & 0.0208 & 6 &10.4  & 9.67 &1.00 & 11.69 & 12.10 & 149 & 254 & 275\\
 J1255$+$04 &193.95403 &  4.30429 & 0.0025 &10 & 4.1  & 9.42 &1.00 & 11.82 & 12.24 & 166 & 279 & 306\\
 J1432$+$06 &217.98907 &  6.25048 & 0.0079 & 5 & 9.8  & 9.55 &0.99 & 11.04 & 11.39 &  89 & 146 & 162\\
 J1535$+$30 &233.81789 & 30.86407 & 0.0061 & 6 &10.4  & 9.23 &1.00 & 11.03 & 11.47 &  71 & 154 & 171\\
 J1548$+$21 &237.17224 & 21.86938 & 0.0072 & 5 &13.0  & 9.45 &0.98 & 11.50 & 11.78 & 142 & 198 & 218\\
 J1801$+$63 &270.45517 & 63.31098 & 0.0259 & 6 &11.7  & 9.71 &1.00 & 11.02 & 11.46 &  71 & 155 & 169\\
 J2332$-$00 &353.10529 & -0.84698 & 0.0176 & 5 &15.6  & 9.45 &1.00 & 11.60 & 12.08 & 139 & 252 & 273\\
 \hline           
 J1049$+$09 &162.50545 &  9.07052 & 0.0334 & 5 & -- & 9.56  & -- & 11.32 & 12.05 & 209 & 231 & 248\\
 J1244$+$62 &191.05040 & 62.24750 & 0.0088 & 5 & -- & 8.78  & -- & 10.48 & 10.78 &  51 &  91 & 102\\
 \hline
\end{tabular}
\tablecomments{1\textwidth}{Columns: (1) Group name. The group
  J1049+09 was identified by \citet{slj+17}, and J1244+62 by
  \citet{psy+24}.  Noticed that the groups J0426$-$04 and J0948$-$03
  have also been reported recently by \citet{psy+26} when we were
  preparing the submission of this paper; (2 - 4) Right ascension,
  declination and spectroscopic redshift of the central galaxy; (5)
  Number of members within a projected radius of 200~kpc and a
  velocity difference of $|\Delta v|\le300 {\rm ~km~s^{-1}}$; (6)
  Overdensity ratio $\delta$ of the number density of galaxies within
  a 200 kpc radius and $|\Delta v|<300~{\rm km~s^{-1}}$ (green boxes
  in Figure~\ref{fig1}) to that within 500 kpc and $|\Delta
  v|<1200~{\rm km~s^{-1}}$ (green and cyan boxes); (7) Total stellar
  mass of all member galaxies; (8) Mass completeness factor, $C_{M*}$,
  derived as the fraction of the estimated stellar mass from
  spectroscopically confirmed member galaxies of the total stellar
  mass of member galaxy candidates also with only photometric data; (9
  - 10) Minimum bound mass and dynamical mass of the group; (11) 3D
  velocity dispersion; (12) Escape velocity at the group's virial
  radius; (13) Estimated virial radius by assuming $M_{\rm vir}=
  M_{\rm dyn} $.}
\end{table}

\begin{figure*}
  \centering
  \includegraphics[angle=0,width=0.56\textwidth]{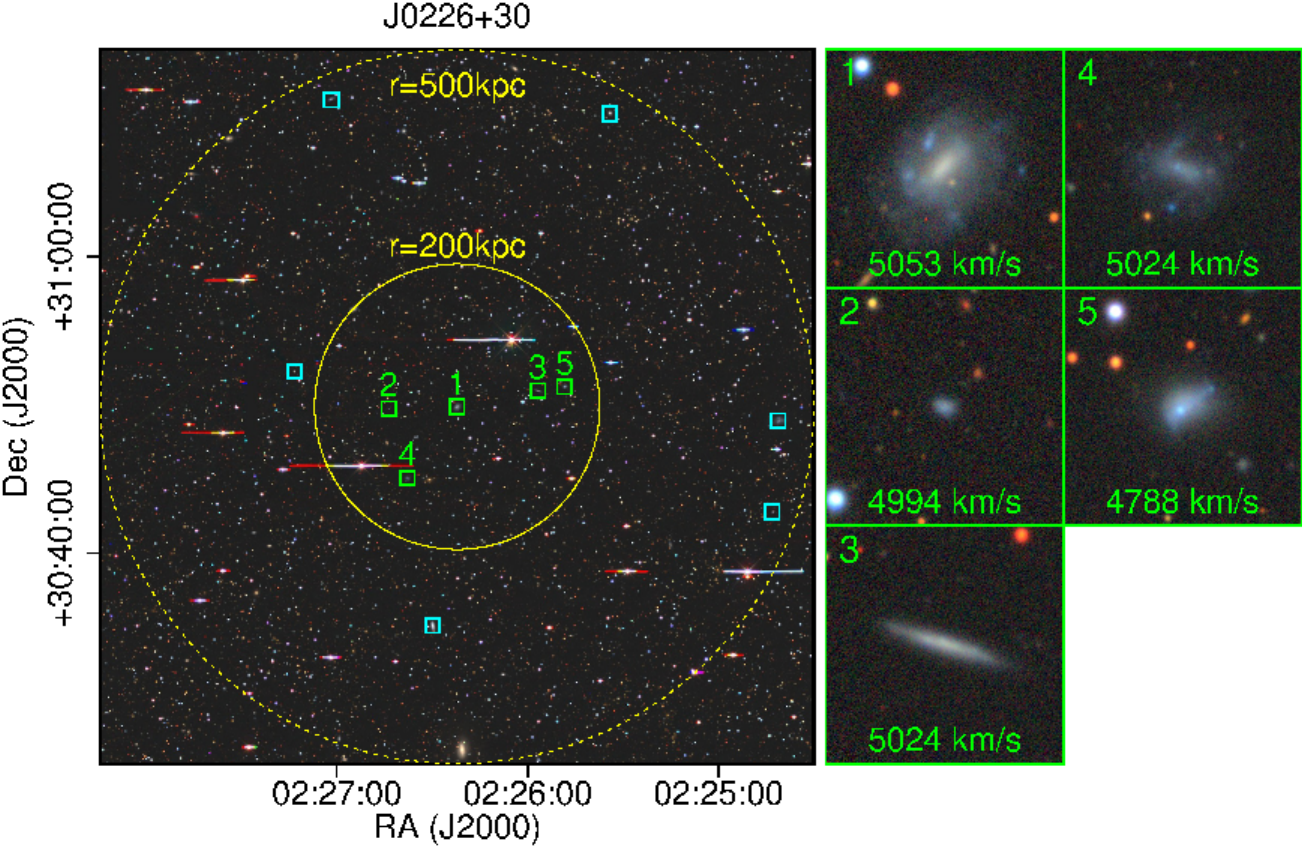} \hspace{0.03\textwidth}
  \includegraphics[angle=0,width=0.342\textwidth]{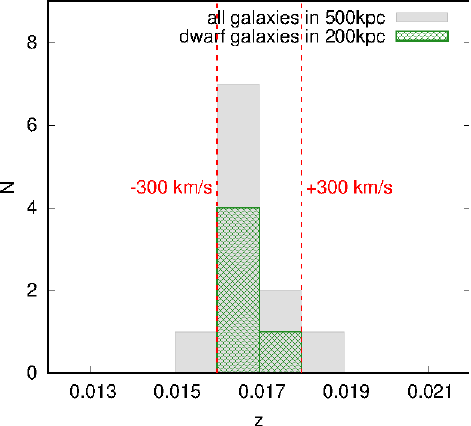}\\
  \caption{Left panel: DESI image of dwarf galaxy group
    J0226+30. Solid and dotted circles denote projected radii of
    200~kpc and 500~kpc, respectively. Cyan boxes mark galaxies within
    500~kpc and $|\Delta v|<$1200~km~s$^{-1}$ relative to the central
    galaxy; green boxes within 200~kpc and $|\Delta
    v|<$300~km~s$^{-1}$ (labeled ``1" - ``5" according to their
    projected distance from the central galaxy) are defined as group
    members. Each green box has a physical scale of 20~kpc, and its
    zoomed-in image is shown on the right, with galaxy IDs and
    line-of-sight velocities labeled. Right panel: Redshift
    distribution of galaxies within 500~kpc and $|\Delta
    v|<$1200~km~s$^{-1}$ relative to the central galaxy (solid
    histograms). Group members within 200~kpc and $|\Delta
    v|<$300~km~s$^{-1}$ are highlighted with hatched
    histograms. Vertical dashed lines denote $\Delta
    v=\pm$300~km~s$^{-1}$ relative to the central galaxy.}
  \label{fig1}
\end{figure*}

\subsection{Search methods}
\label{search}

Clusters or groups of galaxies are generally identified as
over-density regions of galaxy distributions in the three-dimensional
(3D) space, expressed by the 2D projected distribution in the sky
plane plus the radial distribution given by the redshift. The virial
radius and the escape velocity of galaxy groups can be taken as the
basic selection criteria for the group identification. The virial
radius $r_{\rm vir}$ is defined as the spherical radius within which
the total mass can be estimated from the average density $\Delta_{\rm
  vir}(z) \rho_{\rm c}(z)$ by
\begin{equation}
  M_{\rm vir} = \frac{4\pi}{3}r_{\rm vir}^{3} \Delta_{\rm vir}(z)
  \rho_{\rm c}(z).
\end{equation}
Here the virial over-density $\Delta_{\rm vir}(z)$ is taken from
\citet{bn98} as being
\begin{equation}
\Delta_{\rm vir}(z) = 18 \pi^2 + 82\left[\Omega_{m}(z) - 1\right] - 39
\left[\Omega_{m}(z) - 1\right]^2,
\end{equation}
where $\Omega_{m}(z)=\Omega_{m}(1+z)^3H_0^2/H(z)^2$ and
$H(z)=H_0\sqrt{\Omega_{m}(1+z)^3+\Omega_{\Lambda}}$. The critical
density $\rho_{\rm c}(z)$ of the Universe is
\begin{equation}
  \rho_{\rm c}(z)=\frac{3H(z)^2}{8\pi G},
\end{equation}
where $G$ is the gravitational constant.  Assuming the matter
distribution of galaxy groups follows the Navarro-Frenk-White (NFW)
profile \citep{nfw96}, the escape velocity at the group center $v_{\rm
  esc, 0}$ and at the virial radius $v_{\rm esc, vir}$ are given by
\begin{equation}
  \begin{split}
  v_{\rm esc,0}=\sqrt{\frac{2GM_{\rm vir}}{r_{\rm
        vir}}\cdot\frac{c_{\rm vir}}{{\rm ln}(1+c_{\rm vir})-c_{\rm
        vir}/(1+c_{\rm vir})}},\\
  v_{\rm esc,vir}=\sqrt{\frac{2GM_{\rm
        vir}}{r_{\rm vir}}\cdot\frac{{\rm ln}(1+c_{\rm vir})}{{\rm
        ln}(1+c_{\rm vir})-c_{\rm vir}/(1+c_{\rm vir})}},
  \end{split}
\end{equation}
respectively. Here, $c_{\rm vir}$ denotes the concentration parameter
of galaxy groups \citep{dsk+08},
\begin{equation}
c_{\rm vir}(M_{\rm vir},z)=7.85\left(\frac{M_{\rm
    vir}}{2\times10^{12}h^{-1}
  M_{\odot}}\right)^{-0.081}\left(1+z\right)^{-0.71},
\end{equation}
where $h=H_0/100$.

For groups of dwarf galaxies with total masses of $10^{11}~M_{\odot}$
or $10^{12}~M_{\odot}$ at redshift $z=0$, their virial radii $r_{\rm
  vir}$ are approximately 120 or 260~kpc, central escape velocities at
the group center, $v_{\rm esc,0}$, are about 220 or 450 km s$^{-1}$,
respectively.  Referencing the mass ranges ($\lesssim
10^{12}M_{\odot}$) of dwarf galaxy groups identified by \citet{slj+17}
and \citet{psy+24}, we adopted the projected search radius of 200 kpc
and the line-of-sight velocity difference of 300 $\rm km~s^{-1}$ as
being the threshold for identifying dwarf galaxy groups.  We searched
for companions of each dwarf galaxy within a projected radius of
200~kpc and a line-of-sight velocity difference of $|\Delta
v|\le300{\rm~km~s^{-1}}$. We also set the criteria for candidate
groups to host at least five dwarf members. Many redundant candidates
arise from overlapping member galaxies, so we merged groups that share
members and retain unique galaxies only once. To ensure isolation from
massive halos and cluster environments, we rejected any candidate
containing a galaxy with $M_*>10^{10}~M_{\odot}$ within 500~kpc and
$|\Delta v|\le1200{\rm~km~s^{-1}}$ from the brightest member of the
candidate group. After applying these selection criteria, we obtained
139 candidate dwarf galaxy groups.

We then visually inspected the DESI images of all candidates and
removed contaminants, including:
\begin{itemize} 
\item Duplicated observations for one galaxy with a projected
  separation of $>$10~kpc;
\item Massive galaxies with $M_*>10^{10}~M_{\odot}$ that lack
  spectroscopic redshifts, lie within 500 kpc of the candidate group,
  and have a photometric redshift offset $\Delta z_{\rm photo}<0.05$;
\item Massive galaxies misclassified as dwarfs due to underestimated
  masses, such as underestimated distances from peculiar velocities or
  underestimated magnitudes caused by incomplete photometry;
\item Uniformly distributed galaxies without physical concentration.
\end{itemize}
After such validations, we finally obtained 14 new dwarf galaxy
groups, as listed in Table~\ref{tab1}. We do not recover the known
dwarf galaxy groups of J1049$+$09 \citep{slj+17} and J1244$+$62
\citep{psy+24}, since there are only three members of the former with
spectroscopic redshifts in our parent sample, while the latter
contains two closely merging galaxies (separation $<$10~kpc) and our
algorithm retains only one entry for such a pair. Figure~\ref{fig1}
illustrates the projected galaxy distribution (left panel) and
redshift histogram (right panel) for the newly identified dwarf galaxy
group J0226$+$30. Galaxies located within 200~kpc and with $|\Delta
v|\le$300~km~s$^{-1}$ from the central galaxy are classified as group
members, labeled ``1" -- ``5'' according to their projected distance
to the central galaxy. The most massive galaxy in each dwarf galaxy
group is taken as the group center, and labeled as No.1.  The zoom-in
images show that all members are blue galaxies. Table~\ref{tab1}
presents the parameters for all dwarf galaxy groups with at least five
members, including our 14 new systems and two previously known ones.

To assess the spatial concentration degree of the dwarf galaxies in
these groups, we defined the overdensity ratio, $\delta$, by comparing
the galaxy number density $\rho_{200\rm kpc}$ within the projected
radius of 200 kpc and line-of-sight velocity offset $|\Delta
v|<300\rm~km~s^{-1}$ (green boxes in Figure~\ref{fig1} and
Figure~\ref{figA1}) to that $\rho_{500\rm kpc}$ within the projected
radius of 500 kpc and $|\Delta v|<1200\rm~km~s^{-1}$ (green and cyan
boxes in Figure~\ref{fig1} and Figure~\ref{figA1}), i.e.
\begin{equation}
    \delta =  
    \frac{\rho_{200\rm kpc}}{\rho_{500\rm kpc}} =
    \frac{N_{\rm 200kpc}}{V_{\rm 200kpc}} / \frac{N_{\rm 500kpc}}{V_{\rm 500kpc}} 
           = 
           \left(\frac{5}{2}\right)^{3}\frac{N_{\rm 200kpc}}{N_{\rm 500kpc}}.
\end{equation}
For a random or uniform spatial distribution of galaxies, $\rho_{\rm
  200kpc}=\rho_{\rm 500kpc}$, yielding $\delta=1$. The upper limit of
this parameter could come from the case that all galaxies are enclosed
inside the 200 kpc region without any galaxies in the 200 -- 500 kpc
annulus, so that $N_{\rm 200kpc}=N_{\rm 500kpc}$ and
$\delta=15.625$. We listed the $\delta$ values for all newly
identified groups in Table~\ref{tab1}, and one can see that all dwarf
galaxy groups have $\delta>4$.  Here the velocity threshold $|\Delta
v|<1200~\rm km~s^{-1}$ was used for the 500~kpc region for the
$\delta$ estimation.

\begin{figure}
\centering
{\includegraphics[angle=0,width=0.42\textwidth]{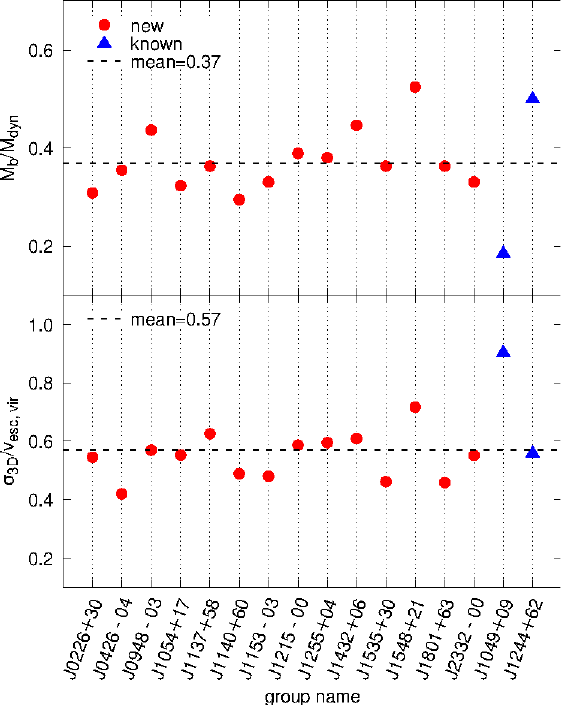}}
\caption{Upper panel: The ratios of the calculated minimum binding
  mass to the dynamical mass ($M_{\rm b}/M_{\rm dyn}$) for dwarf
  galaxy groups, with a mean of 0.37. Circles represent our newly
  identified groups, triangles denote known systems, and the
  horizontal dashed line marks the sample average. Lower panel: The
  ratios of 3D velocity dispersion to virial escape velocity
  ($\sigma_{\rm 3D}/v_{\rm esc,vir}$), with a mean of 0.57.}
\label{fig2}
\end{figure}

All members of dwarf galaxy groups listed in Table A1 have been
identified by using spectroscopic redshifts, which are complete to
some given magnitude limits of available surveys. To further assess
the stellar mass completeness of our groups, we extracted all
photometric galaxies within 200~kpc. The typical error of photometric
redshift is about 0.01, which is much larger than the redshift range
$\Delta z \approx 0.001$ for the velocity threshold of \(\pm300\)
km/s. We only select galaxies whose photometric redshift differs from
the central galaxy’s spectroscopic redshift by less than twice the
photometric redshift uncertainty. This selection yields 155 galaxies;
92 of them are group members with spectroscopic measurements, and the
other 63 galaxies have photometric redshifts only as
candidates. Assuming all these 63 candidates are member galaxies, we
now define the mass completeness factor $C_{M*}$ as the ratio of total
stellar mass from spectroscopically confirmed members to the full
stellar mass, including photometric candidates. We found that the mass
completeness factor $C_{M*}$ reaches nearly 100\% for 13 systems,
except for J0426-04, which is 74\%. This demonstrates that DESI
spectroscopy provides good coverage of faint galaxies in the nearby
universe ($z<0.03$), and that members of dwarf galaxy groups are
roughly complete by means of total stellar mass.

\section{Properties of dwarf galaxy groups}
\label{sect3}

\subsection{Bound status and virial radius of dwarf galaxy groups}

A galaxy group is gravitationally bound if its total mass exceeds the
minimum binding mass, which can be estimated from the velocity
dispersion of member galaxies. The 3D velocity dispersion $\sigma_{\rm
  3D}$ is calculated as \citep[e.g.,][]{slj+17,psy+24}:
\begin{equation}
  \sigma_{\rm 3D}=\sqrt{3}\times\sqrt{\langle v_i^2 \rangle-\langle v_i\rangle^2},
\end{equation}
where $v_i$ denotes the line-of-sight velocity of the $i$-th member
galaxy. A virial equilibrium system satisfies $2T=-U$, with the total
kinetic energy of the system $T\sim\frac{1}{2}M\sigma_{\rm 3D}^2$ and
the gravitational potential energy $U\sim -\frac{GM^2}{\langle
  R_i\rangle}$. The minimum binding mass, $M_{\rm b}$, required for a
system in virial equilibrium is given by
\begin{equation}
  M_{\rm b}=\frac{\sigma_{\rm 3D}^2\times\langle R_i\rangle}{G},
  \label{Mb}
\end{equation}
where $\langle R_i\rangle$ is the mean projected distance of member
galaxies from the group center, and $G$ is the gravitational constant.
For systems not in virial equilibrium, the measured $\sigma_{\rm 3D}$
contains dynamical contributions; therefore, the $M_{\rm b}$
calculated by Eq.(\ref{Mb}) can be regarded as the upper limit of the
true minimum bound mass. The dynamical mass $M_{\rm dyn}$ of the
system is estimated following \citep{htb85}:
\begin{equation}
  \label{dynmass}
  M_{\rm dyn}=\frac{32}{\pi G(N-3/2)}\sum^N_iR_i\Delta v_i^2,
\end{equation}
where $N$ is the number of member galaxies, and $\Delta v_i$ is the
line-of-sight velocity offset of each galaxy relative to the group
center.  We then derived $M_{\rm b}$ and $M_{\rm dyn}$ values for our
14 dwarf galaxy groups and the two known systems, as listed in
Table~\ref{tab1}.  The ratios of $M_{\rm b}/M_{\rm dyn}$ for new
groups (circles) and known systems (triangles) are shown in
Figure~\ref{fig2}. One can see that the dynamical mass of each group
is significantly larger than its minimum binding mass, confirming that
all systems are gravitationally bound.

On the other hand, galaxies are confined within the virial radius if
the system's velocity dispersion is substantially lower than its
escape velocity. Assuming\footnote{Note that this assumption is the
first-order approximation. The dwarf galaxy groups we identified are
dynamically young systems, and may deviate from the dynamical
equilibrium. Nevertheless, our groups are isolated systems with no
recent mergers or interlopers. \citet{fbc17} showed that the dynamical
mass $M_{\rm dyn}$ to the virial mass $M_{\rm vir}$ for both relaxed
and merging systems has a mean offset of only $\sim15\%$.} the
dynamical mass $M_{\rm dyn}$ (Equation~9) is equal to the virial mass
$M_{\rm vir}$, we computed the escape velocities of galaxies at the
virial radius $v_{\rm esc,vir}$ in these dwarf galaxy groups via
Eq.(4) and Eq.(5) assuming the Navarro-Frenk-White density profile
\citep{nfw96} and the mass-concentration relation of
\citet{dsk+08}. The values are listed in Table~\ref{tab1}, and
compared with the velocity dispersion in the lower panel of
Figure~\ref{fig2}. All groups exhibit significantly smaller
$\sigma_{\rm 3D}$ than $v_{\rm esc,vir}$, suggesting that they are
gravitationally bound.

Generally, different density profiles can have significant differences
in the escape velocity within the very central regions
\citep[e.g.,][]{dbm+14}. However, the total masses within $r_{\rm
  vir}$ or $r_{500}$ differ by only a few percent for various density
profiles \citep[e.g.,][]{ml05,ege+19}, causing only small changes in
$v_{\rm esc,vir}$ and not affecting the binding conditions.

\begin{figure}
  \centering
      {\includegraphics[width=0.45\textwidth]{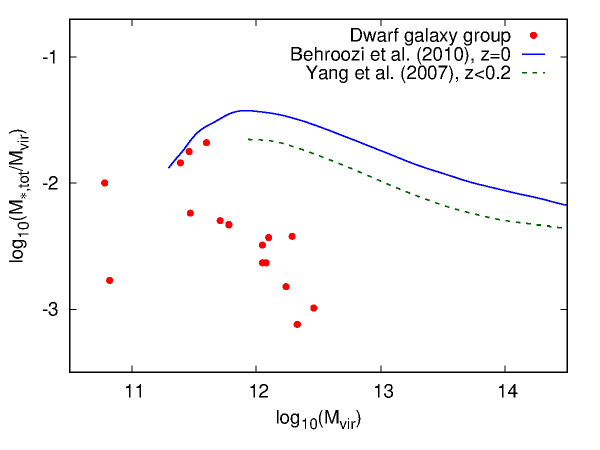}}
      \caption{Stellar mass-halo mass relation for halos with
        different masses. The solid line is the simulation result in
        \citet{bcw10} for galaxies at $z=0$, and the dotted line
        denotes the observational results of \citet{ymv+07} for
        galaxies at $z<0.2$. We adopted $M_{500}=0.63M_{\rm vir}$
        according to fig.9 of \citet{sks+03}. Data for the 16 dwarf
        galaxy groups listed in Table~\ref{tab1} are shown by the red
        dots.}
\label{mr}
\end{figure}

\subsection{Stellar masses of dwarf galaxy groups}

The stellar mass to halo mass relation (SHMR) is a fundamental
diagnostic for probing the cumulative effects of radiative cooling,
star formation, supernova feedback, and environmental quenching across
cosmic structures. Both empirical statistics and hydrodynamical
simulations have established a universal, tightly constrained SHMR
over a broad range of halo masses \citep[e.g.,][]{bcw10, rar20}.

Here we took the estimated virial mass $M_{\rm vir}$ to represent the
halo mass, and found that the total stellar masses of most dwarf
galaxy groups are significantly smaller than expected from the SHMRs
derived in previous studies \citep[e.g.,][]{ymv+07,bcw10,lgb+12},
particularly for systems with halo masses above $10^{12}~M_{\odot}$
(see Figure~\ref{mr}). Our results suggest that these dwarf galaxy
groups, compared to galaxies in massive clusters and field halos, have
a low star formation efficiency in low-mass dark matter halos.

\begin{figure}
  \centering
      {\includegraphics[width=0.4\textwidth]{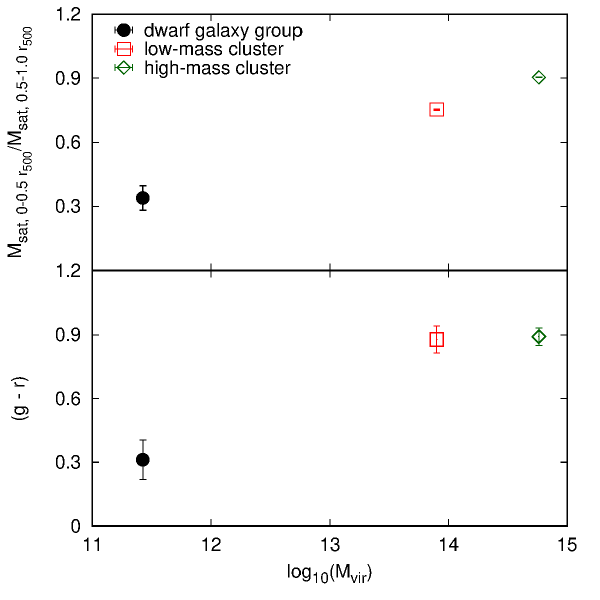}}
      \caption{Comparison of concentration and color of galaxies from
        dwarf galaxy groups, low-mass clusters and high-mass
        clusters. The concentration factor is defined as the ratio of
        the total stellar mass of galaxies within (0 -- 0.5)$r_{500}$
        to that within (0.5 -- 1.0)$r_{500}$. The uncertainties of the
        concentration factor were obtained through error propagation
        based on the stellar mass standard deviations for galaxies in
        the inner and outer regions, respectively, of the stacked
        galaxy samples. The color uncertainties were estimated with
        the median absolute deviation (MAD), a robust method for
        datasets with outliers and non-Gaussian distributions.}
      \label{totalcolor}
\end{figure}

\subsection{Concentration of galaxies and galaxy color}

Young, unrelaxed galaxy systems typically have low concentration of
galaxies because their central gravitational potential is still
growing. In evolved clusters, environmental processes such as
ram-pressure stripping and tidal harassment can remove cold gas from
galaxies, quenching star formation and making galaxies
redder. Meanwhile, galaxies moved from the outskirts toward the center
experience stronger quenching over time, producing a radial color
gradient in which outer galaxies are bluer than inner
galaxies. Concentration and color of member galaxies therefore provide
complementary constraints on the dynamical age of galaxy groups and
clusters.

To compare the properties of galaxies in dwarf galaxy groups and
mature galaxy clusters, we selected three cluster subsamples of
$z<0.05$ from the WH24 catalog \citep{wh24}. Within this redshift
range, the cluster members are complete when $M_*>10^{10}~M_{\odot}$.
We got 122 low-mass clusters with $M_{500} \in
(0.50\pm0.01)\times10^{14}~M_{\odot}$ and 73 massive clusters with
$M_{500}>3.00\times10^{14}~M_{\odot}$. Here $M_{500}$ and $r_{500}$
are directly adopted from the WH24 sample. Since most of the total
stellar mass in these clusters is shown by galaxies with
$M_*>10^{10}~M_{\odot}$ \citep[e.g.,][]{vpa+11}, and the member
galaxies of these clusters are nearly complete above such a mass
threshold within $r_{500}$. We estimated their virial mass and radius
using $M_{500}=0.63M_{\rm vir}$ and $r_{500} = 0.5 r_{\rm vir}$
according to figure 9 of \citet{sks+03}.

We defined the galaxy concentration factor as the ratio of the total
stellar mass of galaxies within (0 -- 0.5) $r_{500}$ to that within
$0.5-1.0~r_{500}$. To reduce statistical biases, we stacked all
galaxies of dwarf galaxy groups and each of the selected cluster
samples to get the average concentration. The results are shown in the
upper panel of Figure~\ref{totalcolor}. Galaxies in dwarf groups
exhibit significantly lower concentration than those in clusters. We
also compared the colors of galaxies between group and cluster
samples. Again, the median $g-r$ color values of the stacked galaxy
populations for all dwarf galaxy groups and each of the cluster
samples were obtained. As shown in the lower panel of
Figure~\ref{totalcolor}, dwarf group member galaxies are considerably
bluer than cluster galaxies. These observed properties of the lower
concentration and globally bluer satellites strongly indicate that the
dwarf galaxy groups are dynamically young systems that have only
recently assembled.

\begin{figure}
  \centering
      {\includegraphics[width=0.4\textwidth]{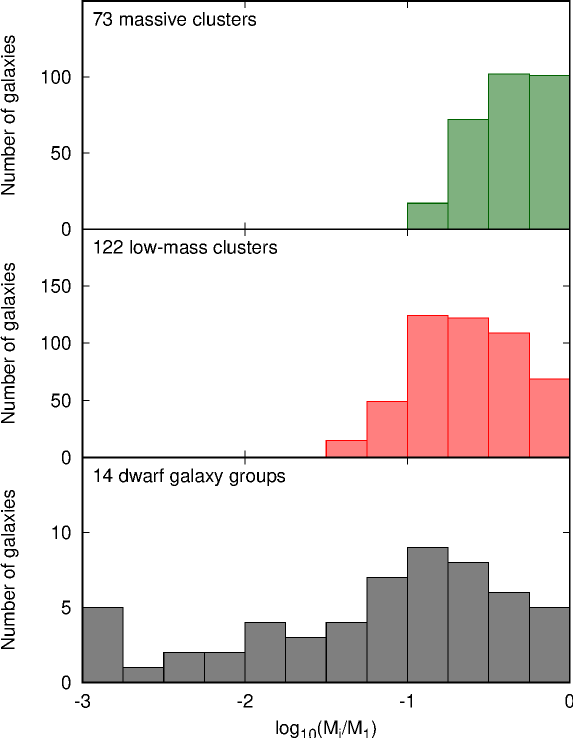}}
      \caption{Distributions of mass ratios $M_{i}/M_1$ of the second
        to fifth massive galaxies ($i=2,3,4,5$) relative to the most
        massive galaxy in 14 dwarf galaxy groups and two samples of
        clusters. The number of groups or clusters used for those
        samples is labeled inside the subpanels.}
      \label{fig4}
\end{figure}

\subsection{Relative mass distribution of member galaxies}

Generally, the brightest galaxy in the central region of a galaxy
group or cluster has mass significantly larger than other member
galaxies. If a cluster is formed through merger events, multiple
massive galaxies formed in different subhalos can merge in and coexist
around the cluster center \citep[e.g.,][]{ggl+22,wjy+26}. Therefore,
the mass distribution of massive galaxies in groups and clusters could
provide insights into their merger history.

We analyzed the mass ratios of the five most massive member galaxies
of 14 dwarf galaxy groups and two samples of clusters selected in
Section 3.3. The mass ratios of the second to the fifth most massive
galaxies are obtained by comparing their masses to that of the most
massive galaxy. We then obtained the distribution of these mass ratios
for the stacked sample of the second to the fifth most massive
galaxies from the sample of dwarf galaxy groups and the two samples of
clusters, as shown in Figure~\ref{fig4}.  Massive galaxy clusters
probably have undergone multiple major mergers during their evolution,
thereby retaining several luminous galaxies from multiple progenitor
subclusters. In contrast, dwarf galaxy groups have not experienced
major mergers and typically possess only one dominant central galaxy,
and the masses of member galaxies have a wide range in the
distribution.

\section{Summary}
\label{sect4}

The discovery and study of dwarf galaxy groups are critical for
understanding galaxy evolution and cosmic structure formation
\citep{slj+17, psy+24, psy+26}. We used spectroscopic datasets of
DESI, SDSS, 2MASS, and also NED to identify 14 new dwarf galaxy groups
with at least five members. All systems are gravitationally bound
through dynamical-to-binding mass comparisons and checks of velocity
dispersion versus virial escape velocity. Dwarf galaxy groups with
$M_{\rm dyn}>10^{12}~M_{\odot}$ show a significantly lower fraction of
stellar mass than that expected by the universal SHMR. We compared the
properties of dwarf galaxy groups and galaxy clusters of different
masses, and found that the dwarf galaxy groups exhibit lower
concentration and bluer colors than clusters. The mass ratio
distributions of the second-to-fifth most massive member galaxies to
the most massive galaxy suggest that dwarf galaxy groups have not
experienced merging processes. We therefore conclude that dwarf galaxy
groups are young, newly formed systems from the dark matter halos in
the local Universe.

\begin{acknowledgements}

We sincerely thank the referee for careful reading and helpful 
comments, which helped us to improve our manuscript.
The authors are supported by the National Natural Science Foundation
of China (Grant Nos 12588202, 12041303), the Chinese Academy of
Sciences via project JZHKYPT-2021-06, Science Research Grants from the
China Manned Space Project (Grant No. CMS-CSST-2025-A04), and the
National SKA Program of China (Grant No. 2022SKA0120103).

The DESI Legacy Imaging Surveys consist of three individual and
complementary projects: the Dark Energy Camera Legacy Survey (DECaLS),
the Beijing-Arizona Sky Survey (BASS), and the Mayall z-band Legacy
Survey (MzLS). DECaLS, BASS, and MzLS include data obtained,
respectively, at the Blanco telescope, Cerro Tololo Inter-American
Observatory, NSF’s NOIRLab; the Bok telescope, Steward Observatory,
University of Arizona; and the Mayall telescope, Kitt Peak National
Observatory, NOIRLab. NOIRLab is operated by the Association of
Universities for Research in Astronomy (AURA) under a cooperative
agreement with the National Science Foundation. Pipeline processing
and analyses of the data were supported by NOIRLab and the Lawrence
Berkeley National Laboratory. The Legacy Surveys also use data
products from the Near-earth Object Wide-field Infrared Survey
Explorer (NEOWISE), a project of the Jet Propulsion
Laboratory/California Institute of Technology, funded by the National
Aeronautics and Space Administration. The Legacy Surveys were
supported by the Director, Office of Science, Office of High Energy
Physics of the U.S. Department of Energy; the National Energy Research
Scientific Computing Center, a DOE Office of Science User Facility;
the U.S. National Science Foundation, Division of Astronomical
Sciences; the National Astronomical Observatories of China; the
Chinese Academy of Sciences; and the Chinese National Natural Science
Foundation. LBNL is managed by the Regents of the University of
California under contract to the U.S. Department of Energy. The
complete acknowledgments can be found at
\url{https://www.legacysurvey.org/acknowledgment/}.

Funding for the Sloan Digital Sky Survey IV has been provided by the
Alfred P. Sloan Foundation, the U.S. Department of Energy Office of
Science, and the Participating Institutions. SDSS-IV acknowledges
support and resources from the Center for High-Performance Computing
at the University of Utah. The SDSS website is \url{www.sdss.org}.

This publication makes use of data products from 2MASS, which is a
joint project of the University of Massachusetts and the Infrared
Processing and Analysis Center/California Institute of Technology,
funded by the National Aeronautics and Space Administration and the
National Science Foundation. This research has made use of the 
NASA/IPAC Extragalactic Database (NED), which is funded by NASA and 
operated by the California Institute of Technology. 

\end{acknowledgements}

\appendix

\section{}
We present the member galaxies of the 14 dwarf galaxy groups
identified in this study in Table~\ref{tabA1}. Member galaxies are
selected to have a stellar mass less than $M_*<10^{9.5}~M_{\odot}$, a
projected distance from the central galaxy less than 200~kpc, and a
line-of-sight velocity difference relative to the central galaxy less
than 300~km~s$^{-1}$. The stellar masses of the central galaxies in
groups J1054+17 and J1137+58 are $M_*=10^{9.561}M_{\odot}$ and
$M_*=10^{9.585}M_{\odot}$, respectively, which slightly exceed the
threshold of $M_*=10^{9.5}M_{\odot}$. Nevertheless, we still include
these two systems in our final sample. 

\begin{table}
\caption[]{Parameters of member galaxies in the newly identified dwarf galaxy groups
\label{tabA1}}
\small
\centering
\setlength{\tabcolsep}{2pt}
\begin{tabular}{ccccccccccc}
\hline
Name & RA & DEC & $v^{a}$ & $m_g$ & $m_r$ & $m_z$ & $W_1$ & $W_2^{b}$ & $\log M_*$ & $D$ \\
     & (J2000) & (J2000) & (km s$^{-1}$) & (mag) & (mag) & (mag) & (mag) & (mag) & ($M_\odot$) & (kpc) \\
(1) & (2) & (3) & (4) & (5) & (6) & (7) & (8) & (9) & (10) & (11)\\
 \hline
J0226$+$30 [1]  & 36.59251 & 30.83102 & 5053 &16.307 &15.799 &15.479 &16.284 &16.996  &8.992 &  0\\
J0226$+$30 [2]  & 36.68203 & 30.82892 & 4994 &19.589 &19.296 &19.137 &20.099 &21.271  &7.212 & 98\\
J0226$+$30 [3]  & 36.48676 & 30.84905 & 5024 &17.633 &17.154 &16.840 &17.221 &17.925  &8.462 &118\\
J0226$+$30 [4]  & 36.65762 & 30.75080 & 5024 &17.381 &17.064 &16.878 &18.192 &18.801  &8.176 &124\\
J0226$+$30 [5]  & 36.45196 & 30.85361 & 4788 &17.289 &16.998 &16.864 &17.630 &18.138  &8.240 &156\\
J0426$-$04 [1]  & 66.62140 & -4.67019 & 3369 &17.934 &17.583 &17.372 &18.409 &19.347  &7.622 &  0\\
J0426$-$04 [2]  & 66.55425 & -4.68136 & 3428 &18.505 &18.179 &18.023 &19.030 & --     &7.325 & 57\\
J0426$-$04 [3]  & 66.56559 & -4.61083 & 3369 &19.513 &19.379 &19.444 &20.023 &20.911  &6.666 & 69\\
J0426$-$04 [4]  & 66.71145 & -4.58387 & 3398 &17.377 &17.323 &17.284 &18.569 &19.929  &7.539 &105\\
J0426$-$04 [5]  & 66.69563 & -4.55140 & 3398 &19.123 &18.985 &18.884 &19.272 &19.888  &7.044 &118\\
J0948$-$03 [1]  & 147.07855 & -3.73392 & 3872 &14.531 &14.208 &13.998 &14.967 &15.516  &9.308 &  0\\
J0948$-$03 [2]  & 147.04176 & -3.71778 & 3990 &18.162 &17.790 &17.573 &18.389 &19.087  &7.752 & 39\\
J0948$-$03 [3]  & 147.12970 & -3.75250 & 4020 &17.122 &16.829 &16.669 &17.502 &18.123  &8.143 & 53\\
J0948$-$03 [4]  & 147.10097 & -3.87140 & 3902 &16.557 &16.332 &16.231 &17.292 &18.121  &8.232 &135\\
J0948$-$03 [5]  & 147.22115 & -3.70273 & 3990 &17.078 &16.768 &16.598 &17.374 &18.085  &8.184 &142\\
J0948$-$03 [6]  & 147.11070 & -3.88935 & 3872 &18.613 &18.408 &18.306 &19.364 &19.760  &7.281 &154\\
J1054$+$17 [1]  & 163.70267 & 17.62122 & 1167 &12.874 &12.174 &11.683 &12.186 &12.779  &9.561 &  0\\
J1054$+$17 [2]  & 163.77766 & 17.46247 & 1167 &16.141 &15.829 &15.673 &16.538 &17.137  &7.412 & 51\\
J1054$+$17 [3]  & 163.56389 & 17.81031 &  958$^a$ &15.704 &15.343 &15.129 &16.276 &17.316  &7.796 & 69\\
J1054$+$17 [4]  & 163.62262 & 17.34385 & 1107 &13.739 &13.092 &12.566 &13.177 &13.605  &9.092 & 84\\
J1054$+$17 [5]  & 163.62943 & 17.28464 & 1107 &12.998 &12.572 &12.243 &12.968 &13.445  &9.019 &100\\
J1054$+$17 [6]  & 163.31058 & 17.84144 & 1137 &17.118 &16.811 &16.630 &17.425 &18.086  &7.015 &126\\
J1054$+$17 [7]  & 163.25043 & 17.57403 & 1137$^a$ &13.389 &12.952 &12.666 &14.135 &14.714  &8.947 &129\\
J1054$+$17 [8]  & 164.16109 & 17.38367 &  928 &17.691 &17.388 &17.244 &18.063 &18.581  &6.544 &145\\
J1054$+$17 [9]  & 163.87016 & 17.14156 & 1077 &18.376 &18.303 &18.472 &18.329 &18.887  &7.747 &147\\
J1054$+$17 [10] & 163.93523 & 17.00507 & 1137 &15.141 &14.722 &14.444 &15.121 &15.725  &8.141 &191\\
J1054$+$17 [11] & 163.08908 & 17.93527 & 1287$^a$ &14.739 &14.065 &13.608 &14.343 &15.003  &8.507 &197\\
J1137$+$58 [1]  & 174.26193 & 58.41549 & 1167 &13.043 &12.285 &11.791 &12.532 &13.126  &9.585 &  0\\
J1137$+$58 [2]  & 174.03290 & 58.44206 & 1346 &20.019 &19.509 &19.354 &20.544 &20.787  &6.668 & 36\\
J1137$+$58 [3]  & 174.11021 & 58.19162 & 1227 &13.939 &13.519 &13.281 &13.976 &14.498  &8.687 & 69\\
J1137$+$58 [4]  & 174.83870 & 58.26871 & 1137 &13.316 &12.819 &13.038 &13.911 &14.475  &8.352 & 98\\
J1137$+$58 [5]  & 174.79190 & 58.60699 & 1376 &20.091 &19.677 &19.487 &20.520 &22.344  &6.061 & 98\\
J1137$+$58 [6]  & 174.53568 & 58.75826 & 1256 &13.482 &12.752 &12.310 &12.810 &13.396  &9.369 &108\\
J1137$+$58 [7]  & 175.02806 & 58.61304 & 1197 &13.683 &13.069 &12.639 &13.110 &13.657  &9.102 &130\\
J1137$+$58 [8]  & 173.82550 & 58.88828 & 1047 &15.919 &15.404 &15.245 &15.946 &16.606  &7.671 &153\\
J1137$+$58 [9]  & 174.64864 & 57.87418 &  958 &15.465 &15.404 &15.584 &16.280 &16.597  &7.381 &169\\
J1140$+$60 [1]  & 175.12971 & 60.29888 & 1227 &13.210 &12.510 &12.072 &13.072 &13.620  &9.459 &  0\\
J1140$+$60 [2]  & 175.11212 & 60.36572 & 1167 &18.504 &18.138 &18.407 &19.107 &19.706  &6.460 & 21\\
J1140$+$60 [3]  & 174.86224 & 60.17254 & 1256 &16.247 &15.815 &16.048 &16.650 &17.476  &7.426 & 56\\
J1140$+$60 [4]  & 174.73425 & 60.61088 & 1167 &18.953 &18.230 &17.766 &18.969 &19.023  &6.673 &113\\
J1140$+$60 [5]  & 175.86263 & 60.67634 & 1286 &15.542 &15.061 &15.126 &15.863 &16.554  &7.679 &160\\
J1153$-$03 [1]  & 178.41873 & -3.99639 & 1585 &13.245 &12.906 &12.716 &13.634 &14.109  &9.011 &  0\\
J1153$-$03 [2]  & 178.13957 & -4.01368 & 1555 &17.750 &17.447 &17.272 &18.076 &18.661  &6.944 &111\\
J1153$-$03 [3]  & 178.13019 & -3.87247 & 1555 &13.360 &13.016 &12.806 &13.714 &14.185  &8.967 &124\\
J1153$-$03 [4]  & 178.53171 & -3.68241 & 1436 &15.794 &15.482 &15.320 &16.083 &16.690  &7.887 &132\\
J1153$-$03 [5]  & 178.39845 & -4.39376 & 1436 &19.763 &19.478 &19.354 &21.105 &21.579  &5.314 &158\\
J1153$-$03 [6]  & 178.13106 & -3.67474 & 1644 &14.412 &14.107 &13.928 &15.206 &15.993  &8.400 &171\\
J1153$-$03 [7]  & 178.24825 & -4.42689 & 1495 &13.994 &13.697 &13.548 &14.523 &15.124  &8.561 &183\\
J1153$-$03 [8]  & 178.02325 & -4.25382 & 1555 &18.505 &18.232 &18.111 &21.268 & --     &6.536 &187\\
J1215$-$00 [1]  & 183.83398 & -0.39807 & 6171 &16.868 &16.216 &15.685 &15.719 &16.206  &9.372 &  0\\
J1215$-$00 [2]  & 183.84251 & -0.36492 & 6406 &20.027 &19.810 &19.700 &20.236 &23.026  &7.284 & 53\\
J1215$-$00 [3]  & 183.79903 & -0.39994 & 6200 &16.951 &16.579 &16.347 &17.125 &17.741  &8.776 & 54\\
J1215$-$00 [4]  & 183.83063 & -0.36114 & 6171 &17.944 &17.869 &17.889 &18.941 &19.622  &7.899 & 58\\
 \hline
\end{tabular}
\tablecomments{1\textwidth}{--- to be continued.}
\end{table}
\addtocounter{table}{-1}
\begin{table}
\caption[]{Parameters of member galaxies in the newly identified dwarf galaxy groups}
\small
\centering
\setlength{\tabcolsep}{2pt}
\begin{tabular}{ccccccccccc}
\hline
Name & RA & DEC & $v^{a}$ & $m_g$ & $m_r$ & $m_z$ & $W_1$ & $W_2^{b}$ & $\log M_*$ & $D$ \\
     & (J2000) & (J2000) & (km s$^{-1}$) & (mag) & (mag) & (mag) & (mag) & (mag) & ($M_\odot$) & (kpc) \\
(1) & (2) & (3) & (4) & (5) & (6) & (7) & (8) & (9) & (10) & (11)\\
 \hline
J1215$-$00 [5]  & 183.86996 & -0.49961 & 6200 &16.798 &16.212 &15.777 &16.178 &16.694  &9.209 &167\\
J1215$-$00 [6]  & 183.79031 & -0.50757 & 6171 &20.574 &20.187 &19.895 &20.254 &20.568  &7.279 &183\\
J1255$+$04 [1]  & 193.95403 &  4.30429 &  749 &12.138 &11.587 &11.124 &11.823 &12.112  &9.343 &  0\\
J1255$+$04 [2]  & 193.94913 &  4.00712 &  689 &19.062 &19.036 &19.543 &20.954 & --     &7.461 & 56\\
J1255$+$04 [3]  & 194.23769 &  4.06470 &  838 &14.805 &14.487 &14.302 &15.170 &15.711  &7.713 & 69\\
J1255$+$04 [4]  & 194.01836 &  3.81277 &  629 &15.820 &15.532 &15.324 &16.857 &18.220  &6.612 & 93\\
J1255$+$04 [5]  & 193.43047 &  4.15411 &  778 &17.081 &16.894 &16.835 &17.638 &18.128  &6.649 &102\\
J1255$+$04 [6]  & 193.41748 &  4.07562 &  898 &17.035 &16.615 &16.378 &17.269 &18.032  &6.951 &109\\
J1255$+$04 [7]  & 193.31070 &  4.46329 &  719 &13.097 &12.760 &12.587 &13.259 &13.733  &8.566 &123\\
J1255$+$04 [8]  & 194.32547 &  4.99126 &  868 &18.952 &18.877 &18.834 &20.161 &19.821  &5.909 &146\\
J1255$+$04 [9]  & 193.06491 &  4.45734 &  689 &17.806 &17.577 &17.401 &18.382 &19.208  &5.460 &168\\
J1255$+$04 [10] & 193.52592 &  3.41016 &  928 &19.082 &18.616 &18.334 &19.684 &20.378  &6.044 &185\\
J1432$+$06 [1]  & 217.98907 &  6.25048 & 2359 &13.899 &13.382 &12.947 &13.371 &13.674  &9.480 &  0\\
J1432$+$06 [2]  & 217.95738 &  6.24100 & 2240 &18.813 &18.553 &18.419 &19.360 &20.103  &6.688 & 19\\
J1432$+$06 [3]  & 218.03419 &  6.18978 & 2389 &15.481 &15.289 &15.177 &16.224 &16.857  &8.198 & 45\\
J1432$+$06 [4]  & 217.93382 &  6.15800 & 2329 &15.322 &14.920 &14.654 &15.368 &15.933  &8.551 & 64\\
J1432$+$06 [5]  & 217.75131 &  6.05520 & 2359 &18.730 &18.618 &18.639 &19.922 &20.259  &6.498 &181\\
J1535$+$30 [1]  & 233.81789 & 30.86407 & 1823 &14.690 &13.956 &13.381 &13.772 &14.230  &9.104 &  0\\
J1535$+$30 [2]  & 233.81200 & 30.70286 & 1734 &18.929 &18.809 &18.966 &20.189 &21.542  &6.017 & 74\\
J1535$+$30 [3]  & 233.68718 & 31.05349 & 1853 &15.629 &15.325 &15.127 &16.149 &16.803  &8.017 &100\\
J1535$+$30 [4]  & 234.07883 & 30.72576 & 1764 &15.661 &15.150 &14.865 &15.599 &16.364  &8.198 &120\\
J1535$+$30 [5]  & 234.08073 & 30.68137 & 1764 &15.025 &14.783 &14.653 &15.504 &16.049  &8.202 &132\\
J1535$+$30 [6]  & 233.93886 & 31.13586 & 1764 &17.907 &17.718 &17.617 &18.811 &19.206  &6.762 &133\\
J1548$+$21 [1]  & 237.17224 & 21.86938 & 2151 &14.153 &13.323 &12.343 &14.906 &15.045  &9.401 &  0\\
J1548$+$21 [2]  & 237.21486 & 21.85550 & 2240 &18.174 &17.983 &18.085 &18.908 &19.674  &6.766 & 23\\
J1548$+$21 [3]  & 237.23108 & 21.75901 & 2002 &16.549 &16.325 &16.223 &17.198 &17.656  &7.556 & 66\\
J1548$+$21 [4]  & 237.31911 & 21.83179 & 2091 &15.342 &14.949 &14.688 &15.376 &15.957  &8.430 & 76\\
J1548$+$21 [5]  & 236.92294 & 21.63944 & 2180 &18.643 &18.519 &18.469 &19.487 &20.060  &6.584 &176\\
J1801$+$63 [1]  & 270.45517 & 63.31098 & 7664 &16.082 &15.721 &15.491 &16.100 &16.653  &9.416 &  0\\
J1801$+$63 [2]  & 270.43555 & 63.31669 & 7693 &16.318 &15.946 &15.730 &16.180 &16.658  &9.342 & 20\\
J1801$+$63 [3]  & 270.53360 & 63.29831 & 7606 &19.286 &18.982 &18.867 &19.580 &20.017  &7.800 & 72\\
J1801$+$63 [4]  & 270.31821 & 63.32306 & 7723 &19.012 &18.681 &18.556 &19.296 &20.202  &7.954 &121\\
J1801$+$63 [5]  & 270.51105 & 63.25151 & 7693 &19.867 &19.492 &19.313 &20.072 &20.872  &7.624 &125\\
J1801$+$63 [6]  & 270.49954 & 63.40136 & 7723 &18.645 &18.317 &18.225 &19.082 &19.375  &8.063 &179\\
J2332$-$00 [1]  & 353.10529 & -0.84698 & 5230 &15.777 &15.473 &15.285 &16.012 &16.612  &9.076 &  0\\
J2332$-$00 [2]  & 353.07153 & -0.79774 & 5259 &17.132 &16.679 &16.356 &17.041 &17.618  &8.662 & 79\\
J2332$-$00 [3]  & 353.17142 & -0.75722 & 5318 &17.244 &16.933 &16.755 &17.495 &18.089  &8.417 &147\\
J2332$-$00 [4]  & 352.99899 & -0.80524 & 5259 &16.290 &16.017 &15.862 &16.768 &17.280  &8.764 &150\\
J2332$-$00 [5]  & 353.14935 & -0.73730 & 5083 &17.154 &16.740 &16.474 &17.356 &18.003  &8.504 &155\\
 \hline
\end{tabular}
\tablecomments{1\textwidth}{Columns: (1) Name of the member galaxy; (2 - 3) Right ascension and declination; (4) Line-of-sight velocity; (5 - 9) Apparent magnitudes in the $g$, $r$, $z$, $W_1$ and $W_2$ bands; (10) Estimated stellar mass; (9) Projected distance to the most massive member galaxy.\\
$^a$: The velocity calculated from the spectroscopic redshift from NED.\\
$^b$: Unavailable magnitudes are denoted with ``--''.}
\end{table}

\begin{figure*}
  \includegraphics[angle=0,height=0.33\textwidth]{ms2026-0431_J0226+30.pdf}\\
  \includegraphics[angle=0,height=0.33\textwidth]{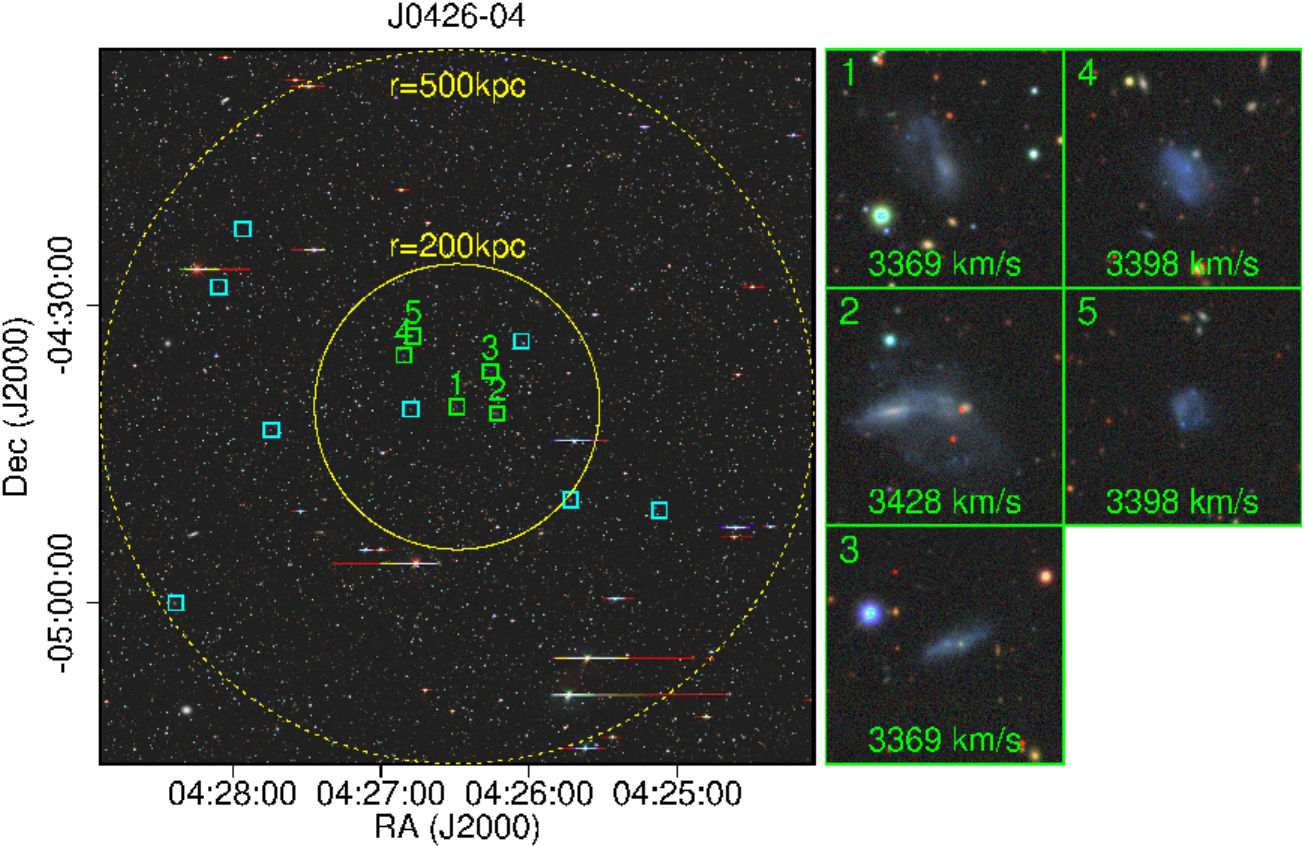}
  \includegraphics[angle=0,height=0.33\textwidth]{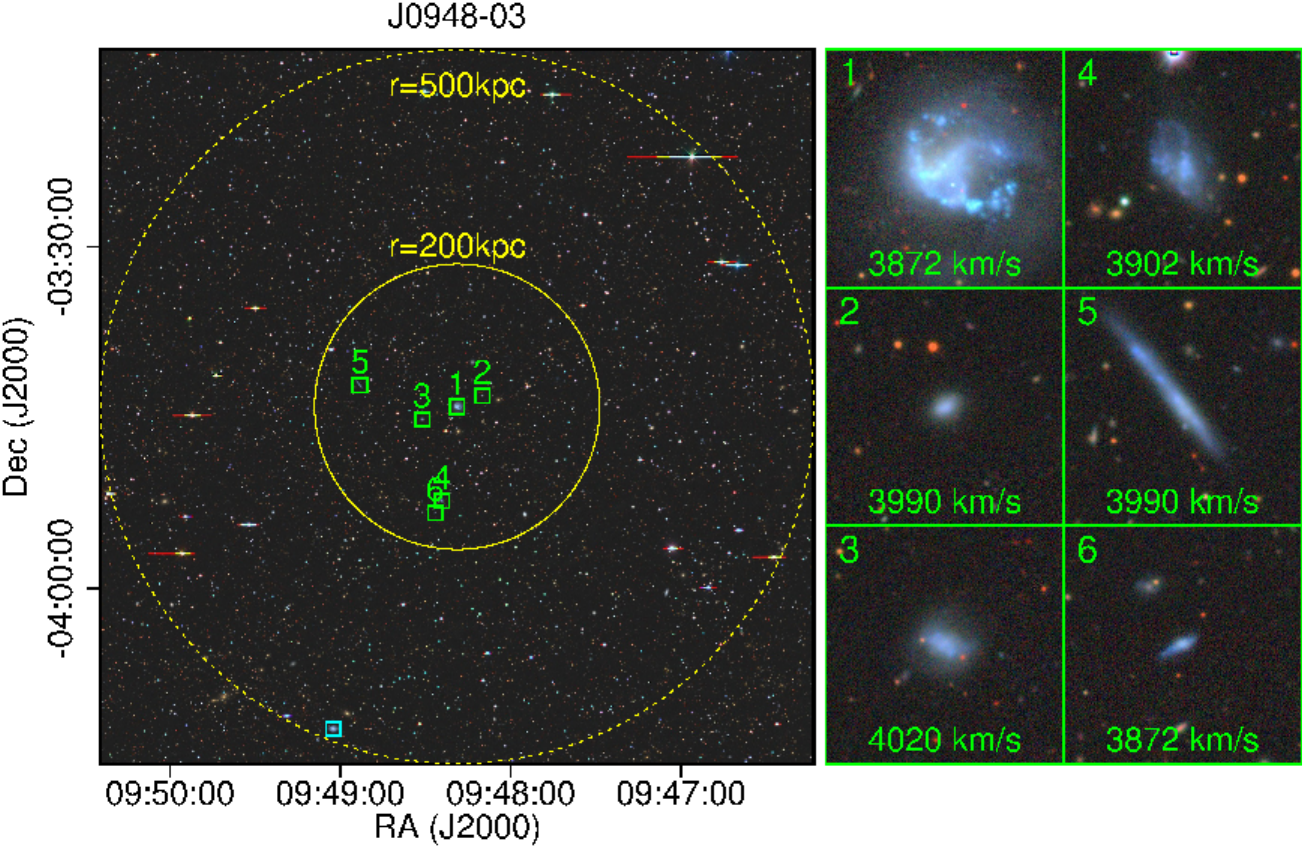}\\
  \includegraphics[angle=0,height=0.34\textwidth]{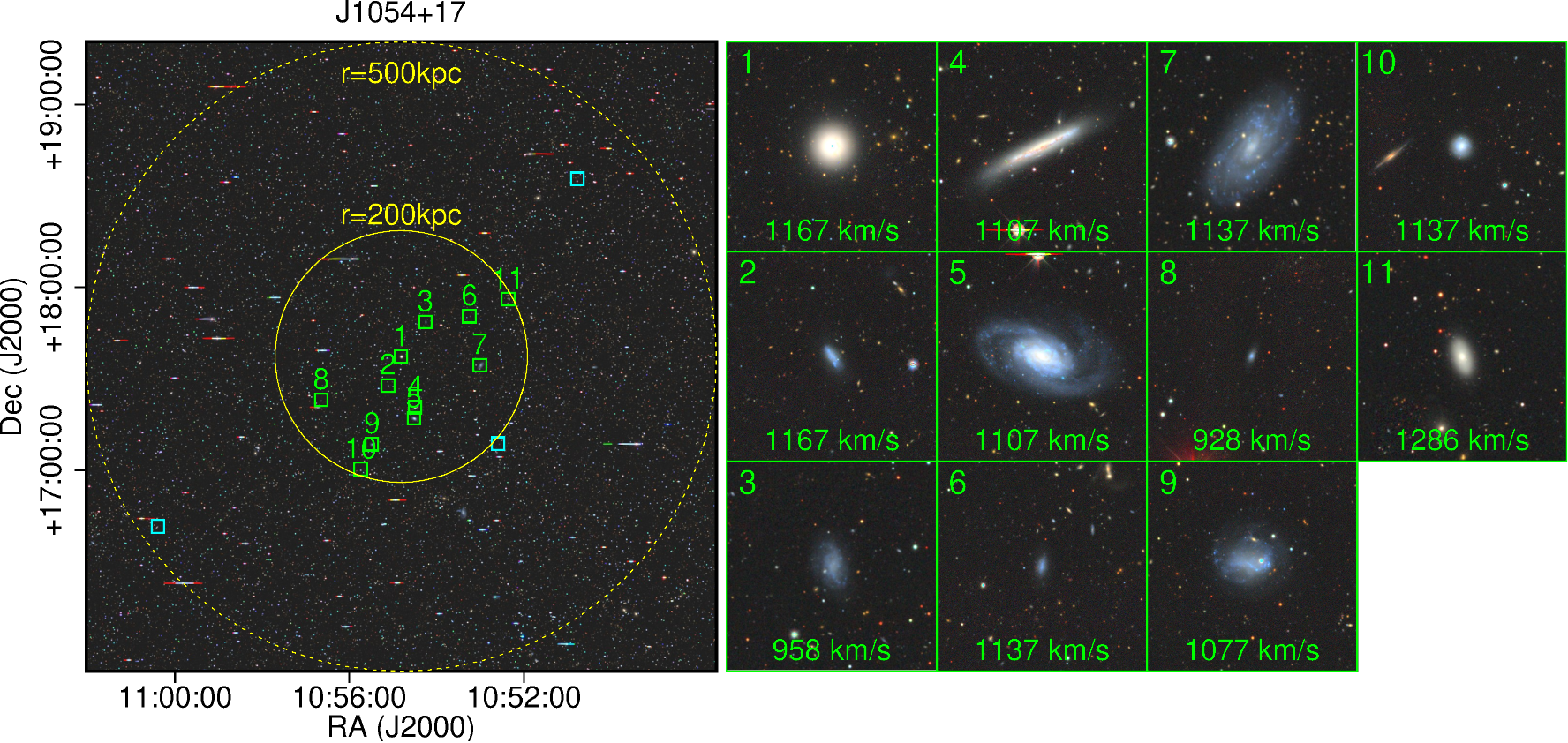}\\
  \includegraphics[angle=0,height=0.34\textwidth]{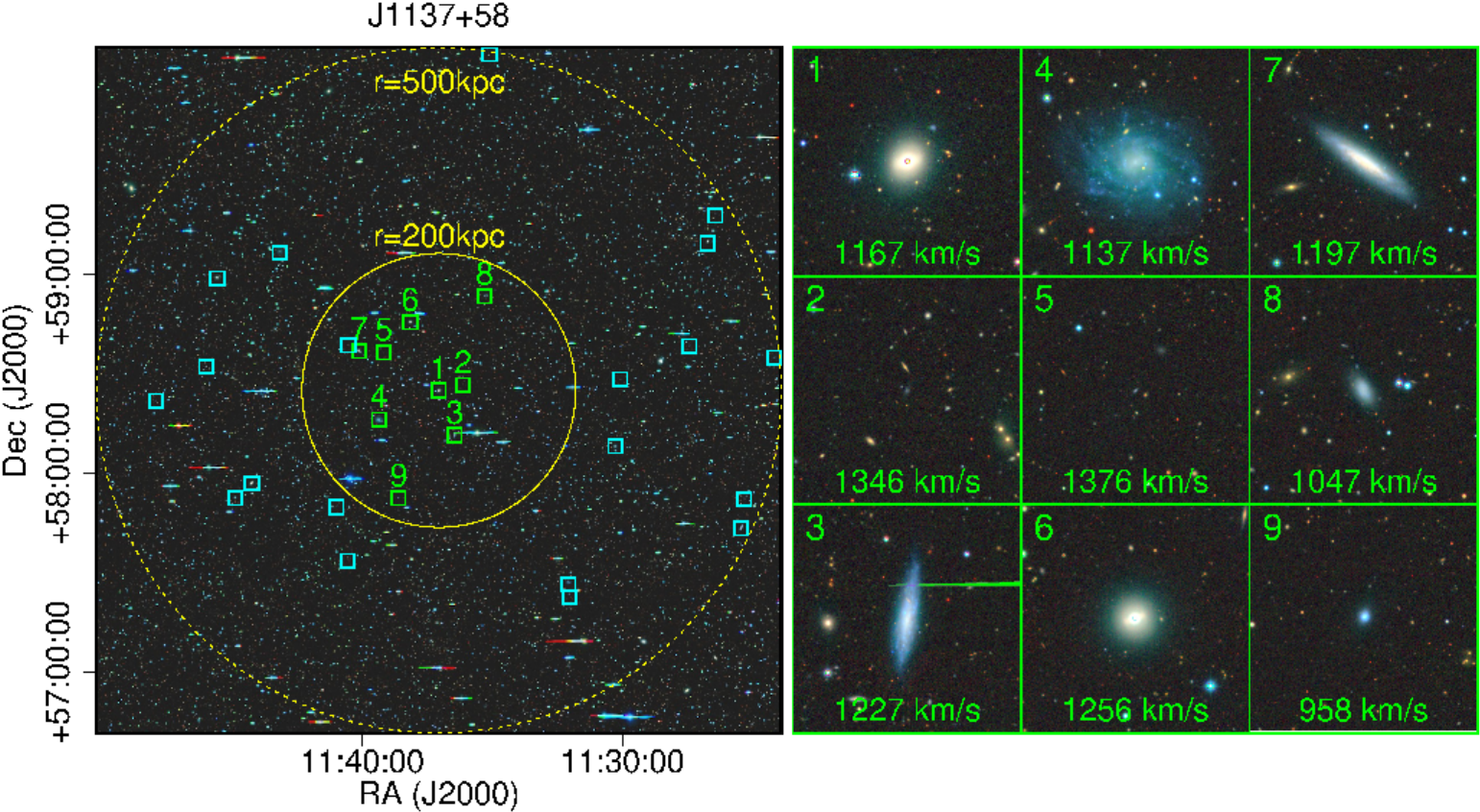} 
  \caption{DESI images of 14 identified dwarf galaxy groups. Symbols in the images have the same meaning as those in the left panel of Figure~\ref{fig1} {--- to be continued.}}
   \label{figA1}
\end{figure*}

\addtocounter{figure}{-1}
 \begin{figure*}
  \includegraphics[angle=0,height=0.33\textwidth]{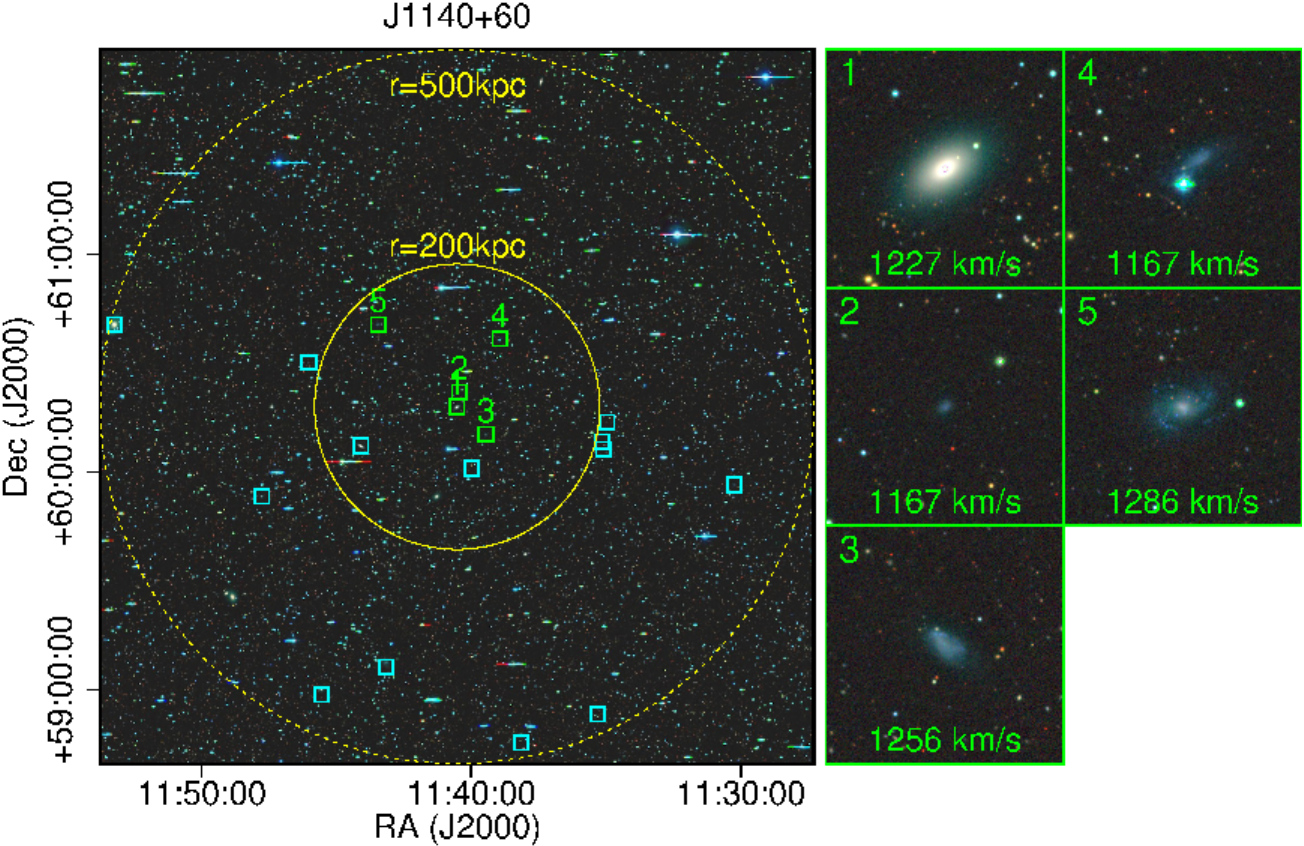}\\
  \includegraphics[angle=0,height=0.33\textwidth]{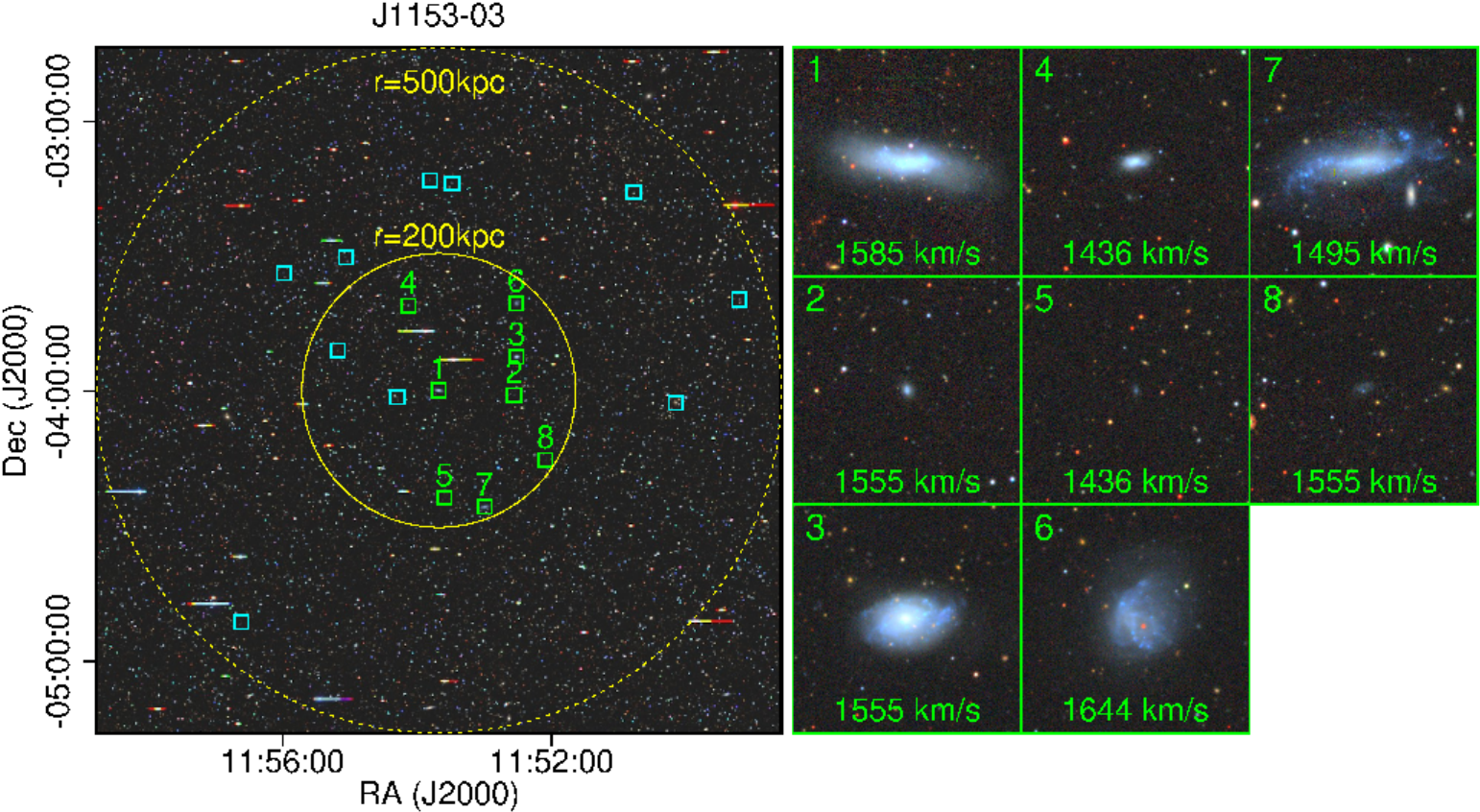}\\
  \includegraphics[angle=0,height=0.33\textwidth]{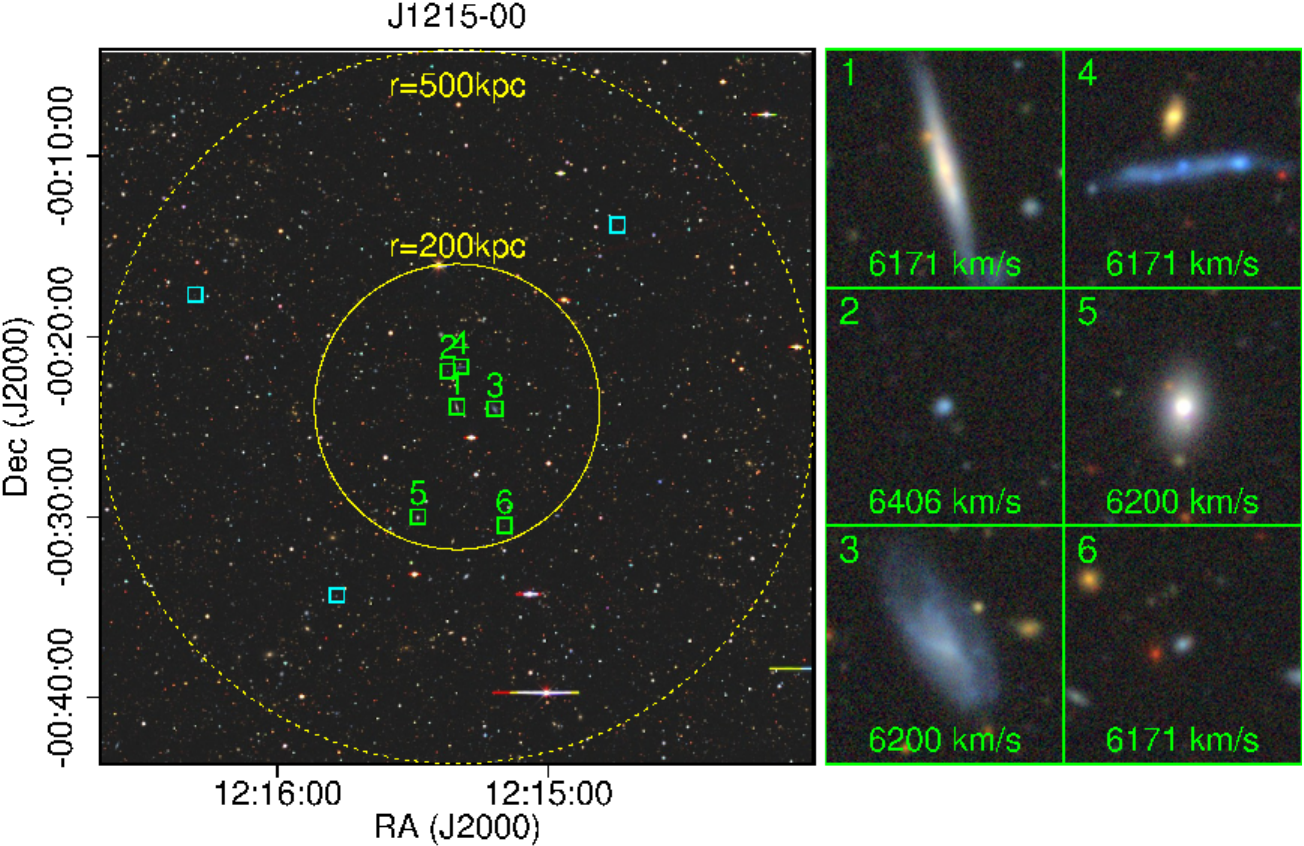}\\
  \includegraphics[angle=0,height=0.33\textwidth]{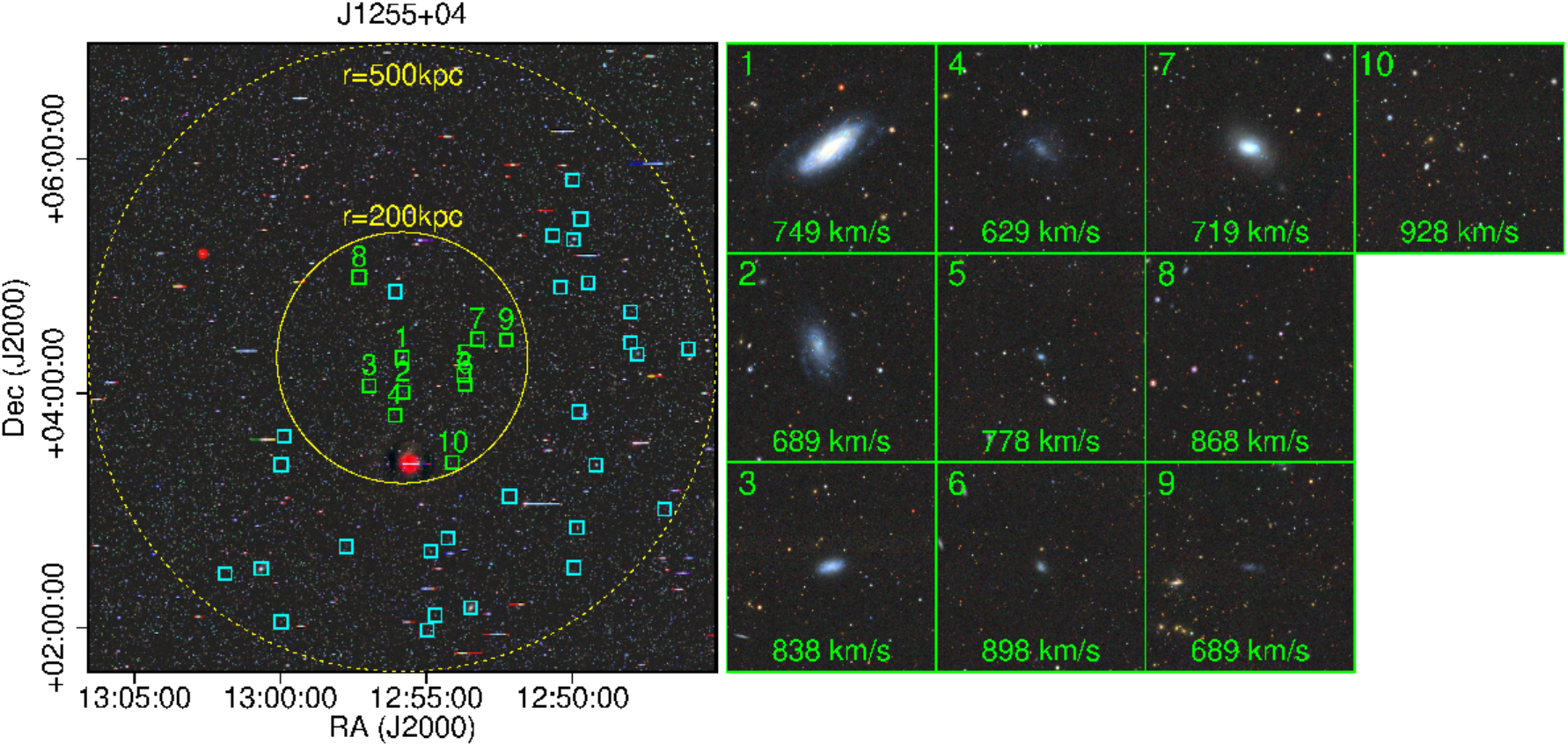}\\
  \caption{{--- to be continued.}}
\end{figure*}
 
\addtocounter{figure}{-1}
\begin{figure*}
  \includegraphics[angle=0,height=0.33\textwidth]{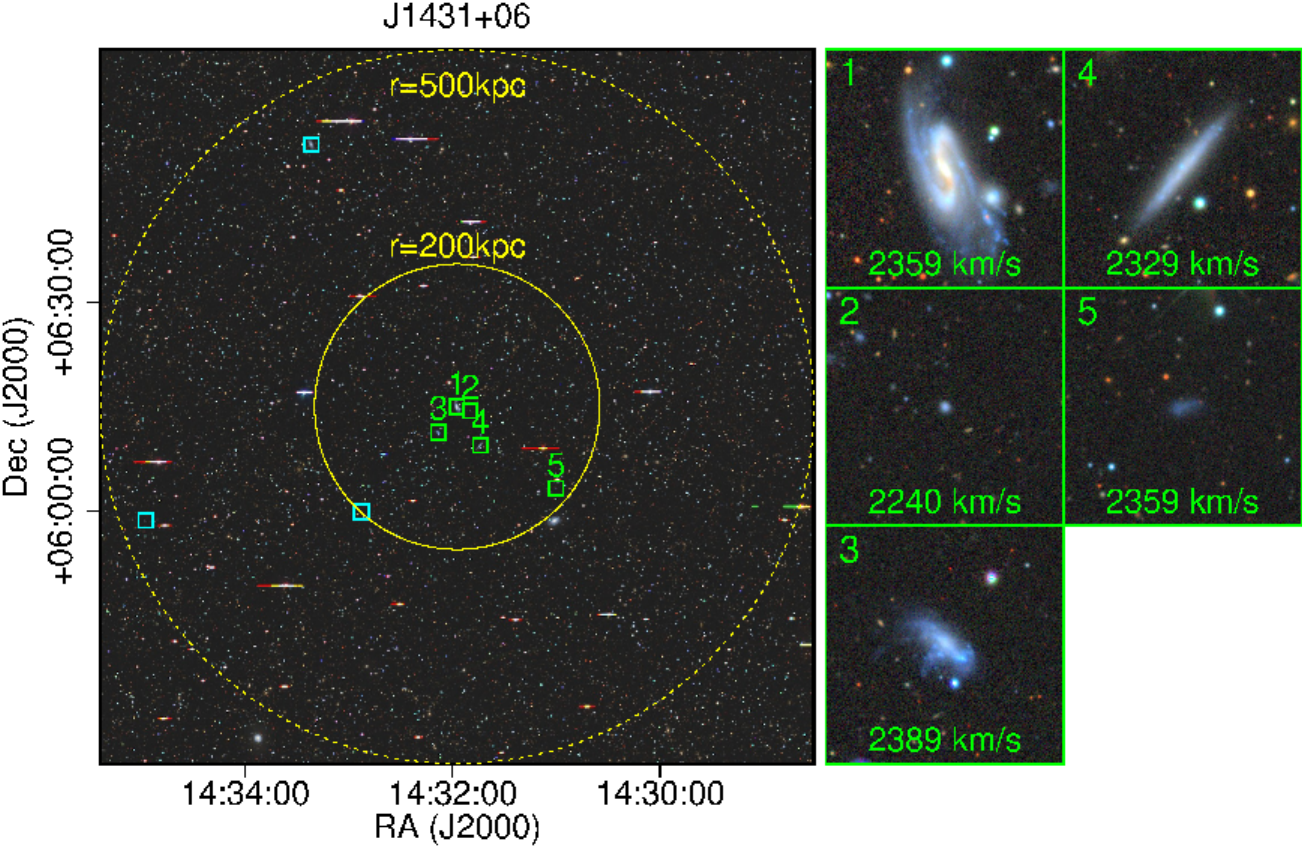}
  \includegraphics[angle=0,height=0.33\textwidth]{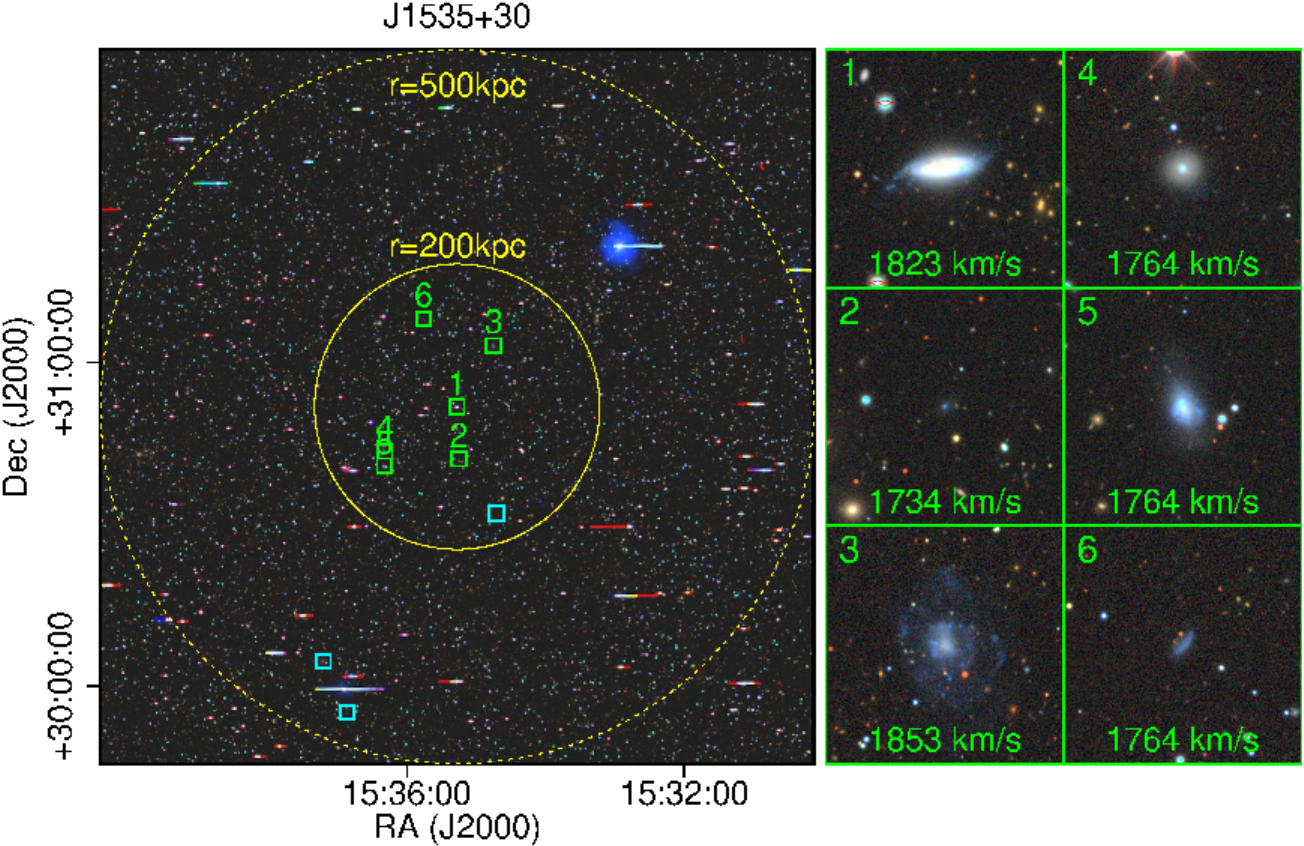}
  \includegraphics[angle=0,height=0.33\textwidth]{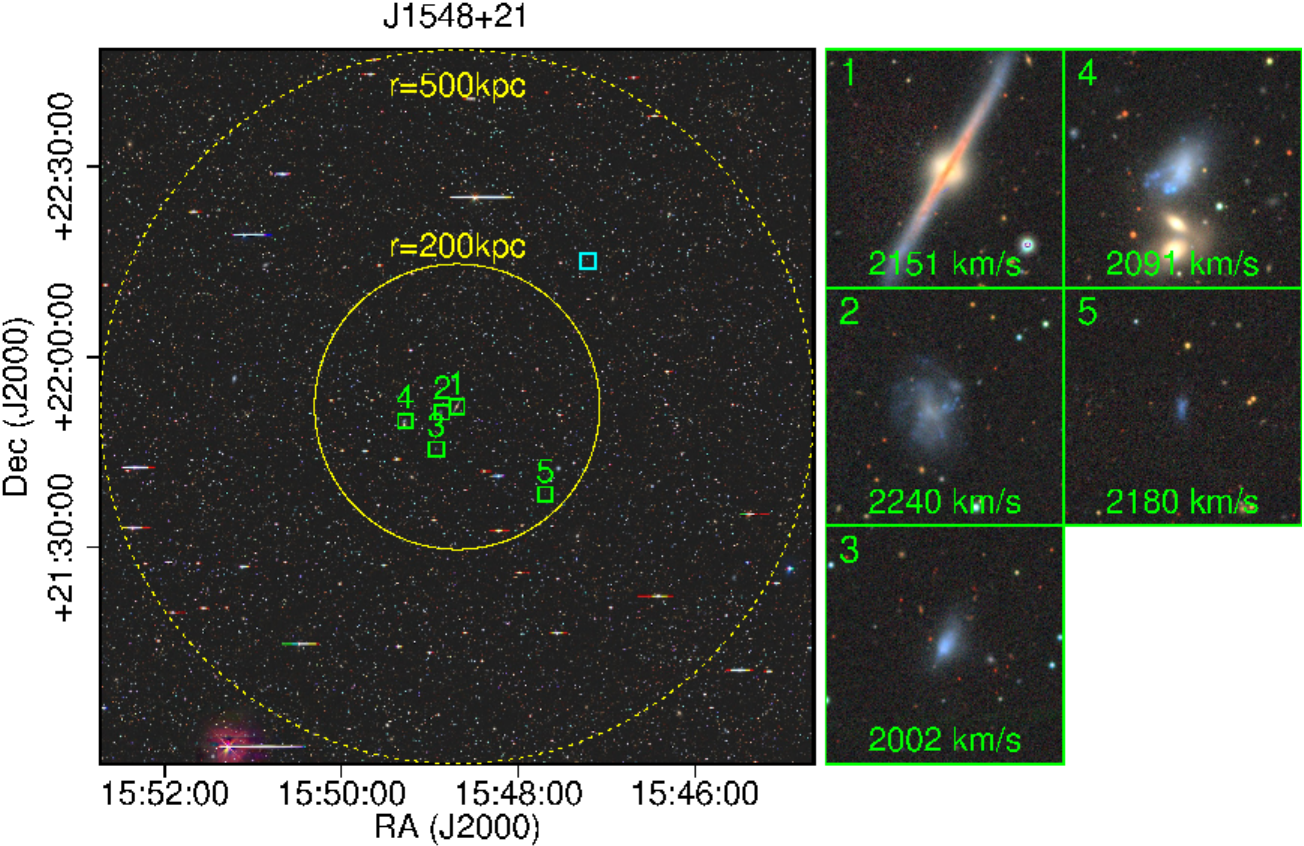}
  \includegraphics[angle=0,height=0.33\textwidth]{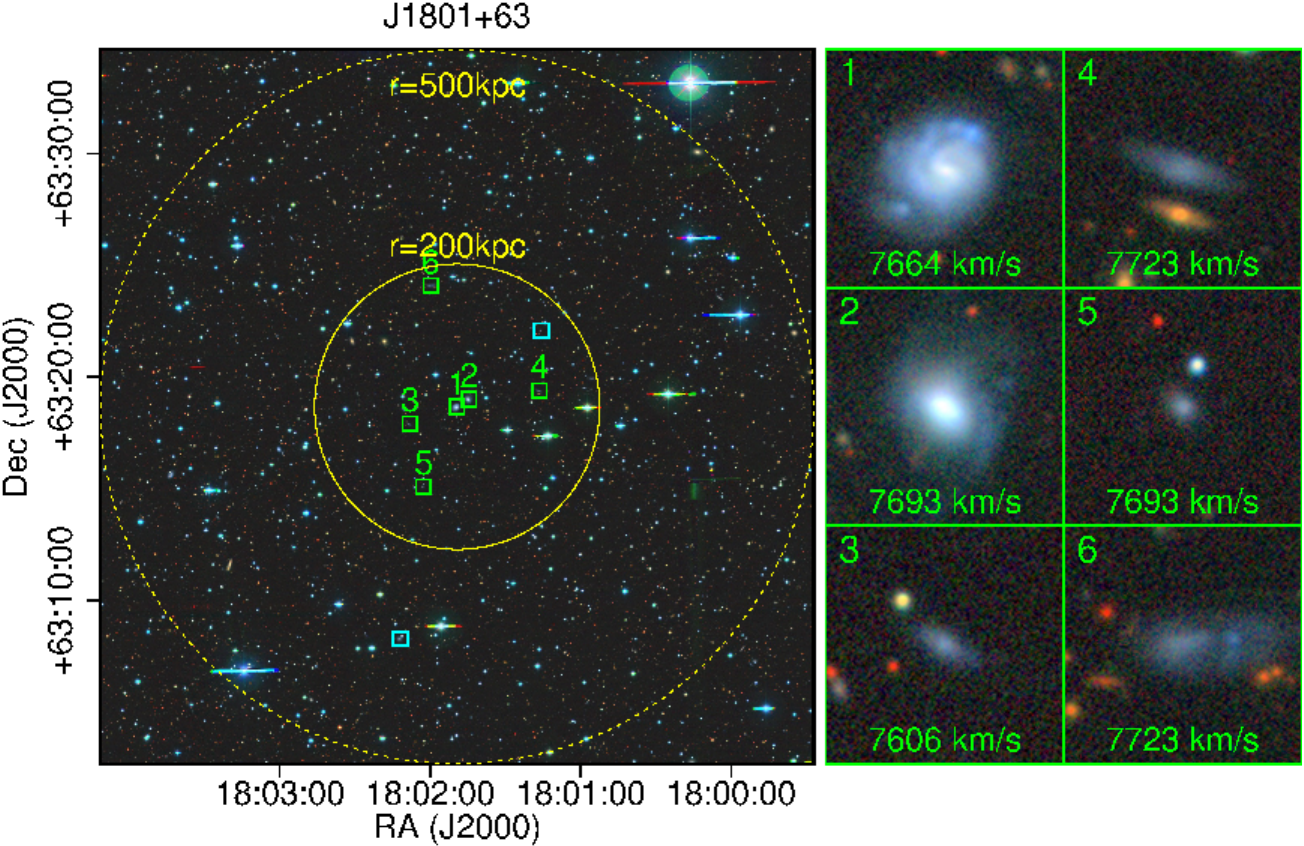}
  \includegraphics[angle=0,height=0.33\textwidth]{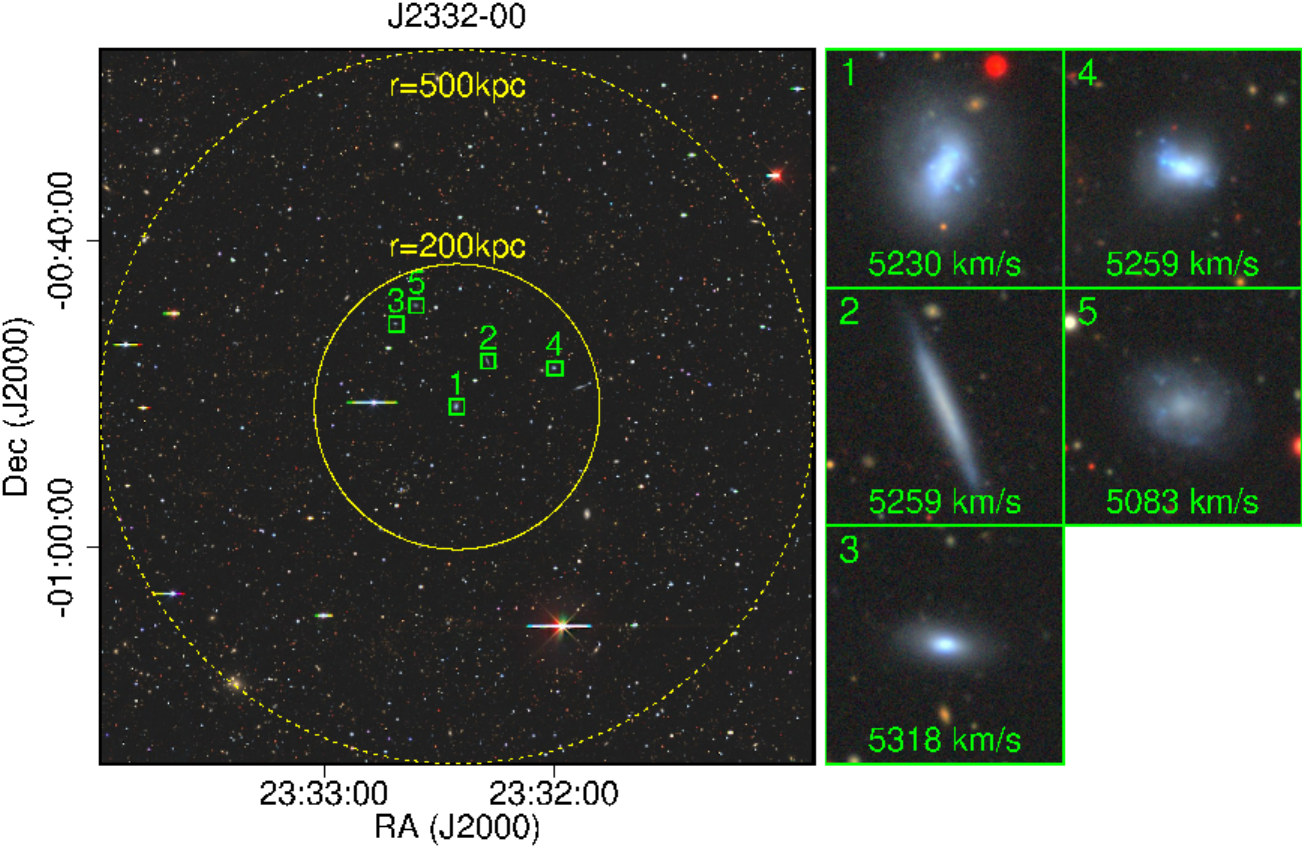}
  \caption{{--- ended.}}
\end{figure*}

\bibliographystyle{raa}
\bibliography{ms2026-0431_ref}{}

\label{lastpage}

\end{document}